\pdfoutput=1
\documentclass[twocolumn]{aastex631}
\usepackage{epstopdf}
\usepackage{xspace}
\usepackage{graphicx}
\usepackage{amssymb}
\usepackage{amsmath}
\usepackage{natbib}
\usepackage{float}
\usepackage{hyperref}

\shorttitle{MIR Features in Dark Globule DC\,314.8--5.1}
\shortauthors{Kosmaczewski et al.}

\begin{document}

\title{Spectroscopic Diagnostics of the Mid-Infrared Features of the Dark Globule, DC\,314.8--5.1,\\ with the Spitzer Space Telescope}

\correspondingauthor{E.~Kosmaczewski}
\email{emily@oa.uj.edu.pl}

\author{E.~Kosmaczewski}
\affiliation{Astronomical Observatory of the Jagiellonian University, ul. Orla 171, 30-244 Krak\'ow, Poland}

\author{{\L}.~Stawarz}
\affiliation{Astronomical Observatory of the Jagiellonian University, ul. Orla 171, 30-244 Krak\'ow, Poland}

\author{W.~R.~M.~Rocha}
\affiliation{Laboratory for Astrophysics, Leiden Observatory, Leiden University, P.O. Box 9513, NL 2300 RA Leiden, The Netherlands}

\author{S.~S.~Shenoy}
\affiliation{Space Science Institute, 4765 Walnut St., Room 203, Boulder, CO 80301}

\author{A.~Karska}
\affiliation{Institute of Astronomy, Faculty of Physics, Astronomy and Informatics, Nicolaus Copernicus University, Grudziadzka 5, 87-100 Toru\'n, Poland}

\begin{abstract}
We present an analysis of the mid-infrared spectra, obtained from the Spitzer Space Telescope, of the dark globule, DC\,314.8--5.1, which is at the onset of low-mass star formation. The target has a serendipitous association with a B-type field star, which illuminates a reflection nebula in the cloud. We focus on the polycyclic aromatic hydrocarbon (PAH) emission features prevalent throughout the mid-infrared range. The analysis of the spectra with the PAHFIT software as well as pypahdb package, shows that (i) the intensities of PAH features decrease over distance from the ionizing star toward the cloud center, some however showing a saturation at larger distances; (ii) the relative intensities of the 6.2 and 8.6 features with respect to the 11.2\,$\mu$m feature remain high throughout the globule,  suggesting a larger cation-to-neutral PAH ratio of the order of unity; the breakdown from pypahdb confirms a high ionized fraction within the cloud; (iii) the pypahdb results display a decrease in large PAH fraction with increased distance from HD\,130079, as well as a statistically significant correlation between the large size fraction and the ionized fraction across the globule; (iv)  the 7.7 PAH feature displays a peak nearer to 7.8\,$\mu$m, suggesting a chemically processed PAH population with a small fraction of UV-processed PAHs; (v) the H$_2$\,S(0) line is detected at larger distances from the ionizing star. All in all, our results suggest divergent physical conditions within the quiescent cloud DC\,314.8--5.1 as compared to molecular clouds with ongoing starformation.
\end{abstract}

\section{Introduction} 
\label{sec:intro}

Polycyclic Aromatic Hydrocarbons (PAHs), containing tens to hundreds of carbon atoms, are an abundant and ubiquitous component of the interstellar medium (ISM); although they are found in many different systems, ranging from molecular clouds in the Milky Way to distant active galaxy nuclei, their origin and distribution are not fully understood \citep[see, e.g., reviews by][]{Tielens08,Joblin11,Li20}. PAHs can be observed via their infrared (IR) fluorescence through the vibrational modes, manifesting in particular in broad emission features at 3.3, 6.2, 7.7, 8.6, 11.3, and 12.7\,$\mu$m. This, however, heavily relies on ultraviolet (UV) radiation for excitation, and therefore quiescent properties of PAHs remain elusive. 

\begin{figure*}[!th]
    \centering
    \includegraphics[width=0.9\textwidth]{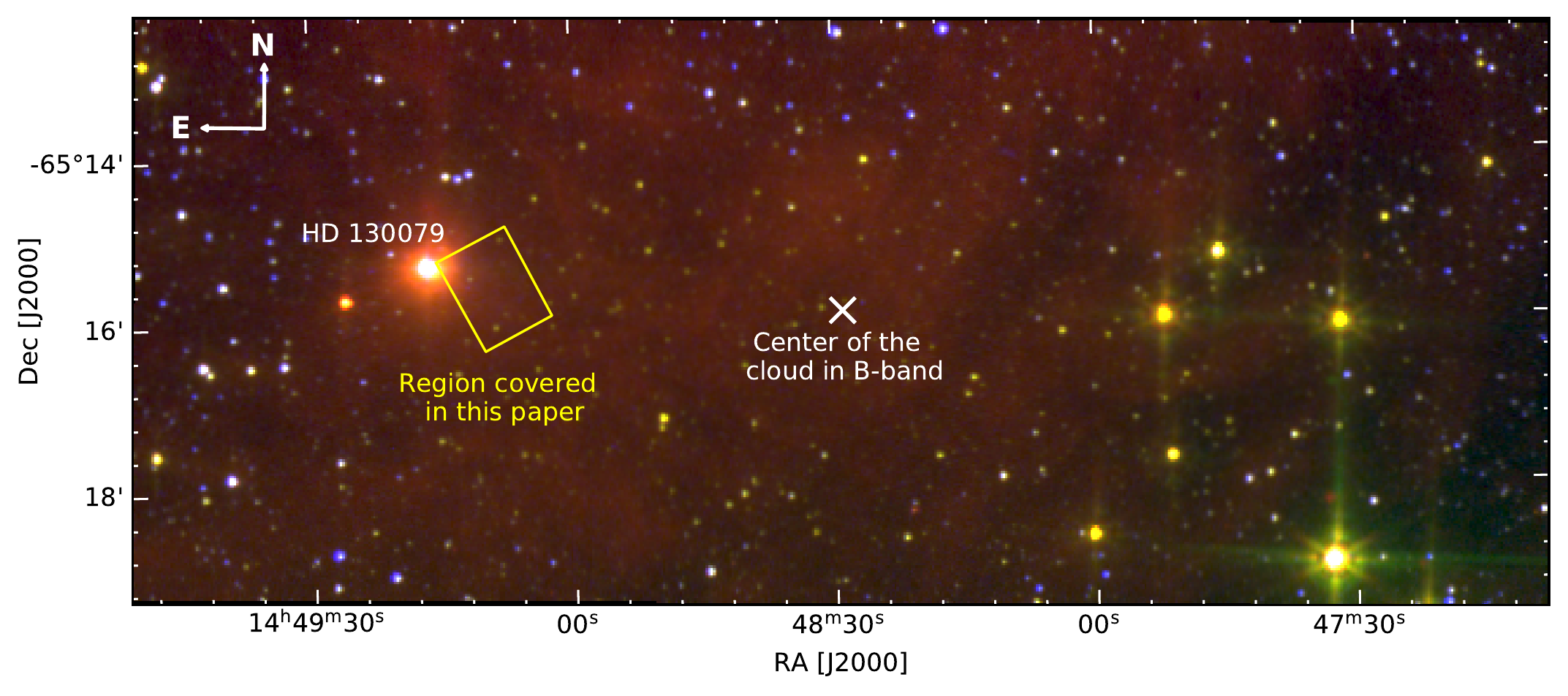}
    \caption{RGB image with a horizontal angular size of $\sim$17$^\prime$ and a vertical angular size of $\sim$7$^\prime$, centered on the dark globule, DC\,314.8--5.1. The red, green and blue colors are taken from Spitzer/IRAC (8.0\,$\mu$m, and 5.8\,$\mu$m), and DSS1 blue band, respectively. The images are combined utilizing the APLpy package v2.0.2 \citep{aplpy2012,aplpy2019}. HD\,130079 can be identified as the bright star located left of center in the eastern boundary of the globule. The globule is the diffuse region in the center, obscuring and reddening the background starlight.}
    \label{DSS}
\end{figure*}

There has been extensive modeling of PAHs in various environments, attempting to disentangle contributions of molecules with different sizes, charge states, or even structure and content, to the IR radiative output at different wavelengths.  It was shown that the 3.3 and 11.3\,$\mu$m features are predominantly due to neutral PAHs, while the 6.2, 7.7, and 8.6\,$\mu$m features are dominated by ionized PAHs; relative intensity of the 6.2 and 7.7\,$\mu$m features, on the other hand, were argued to be sensitive to the number of carbon atoms in a molecule \citep[e.g.,][]{Draine01}. Hence, the intensity ratio diagram ``6.2/7.7 versus 11.3/7.7'' was widely considered as a diagnostic tool enabling constraints on the distributions of PAH sizes and ionization states in various astrophysical sources, especially in ones that can be spatially resolved by IR spectrometers \citep[see][]{Berne07,Visser07,Pineda07,Rosenberg11,Peeters17,Boersma18}. More recently, however, other intensity ratios, such as 11.2/7.7 and 11.2/3.3, have been considered as more reliable proxies for the main characteristics of the PAH population, with the 3.3 feature being particularly sensitive to the number of carbon atoms in a molecule \citep{Maragkoudakis20}.

The key factors determining the plasma ionization, in addition to the amount of the exciting emission, are the gas electron density and temperature. At the same time, the size distribution of PAHs is expected to be shaped by the ionizing continuum, as well as shocks possibly present in a system \citep{Boersma14, Li20, Verstraete11, Siebenmorgen10}. As a result, in star-forming regions --- which are the typical targets for resolved IR spectroscopy  --- PAHs undergo significant processing at early stages of the systems' evolution, and so there is an observational bias in regards to probing the dust and PAH conditions in these systems following the onset of star formation.

\begin{figure}[!ht]
    \centering
    \includegraphics[width=\columnwidth]{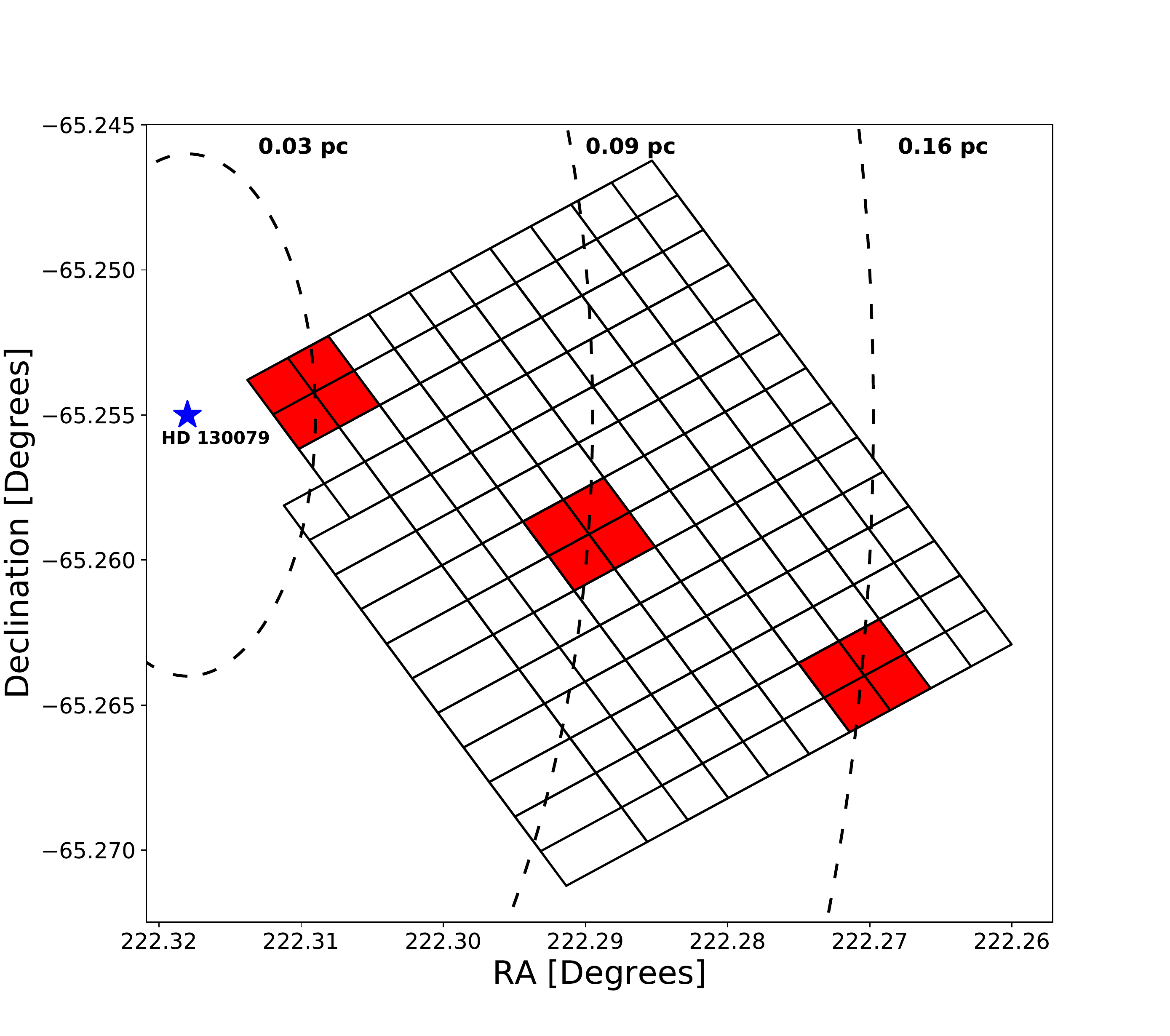}
    \caption{117 overlapping Spitzer IRSMAP MIR spectroscopic regions extracted from the eastern boundary of DC\,314.8--5.1, with increasing distance from HD\,130079. Extraction regions are overlapping to better account for subtle changes in features over distance.  Red shaded regions mark three of the extraction regions which correspond to the spectra given in Figure\,\ref{Lowres}, with respective distance arcs marked by dashed lines. Each cube spans $2 \times 2$ pixels with an angular width of $\sim 5^{\prime\prime}$ per pixel, which at a distance of $435 \pm 4$\,pc (a systematic error of less than 1\%) translates to a spatial size of a box width of $\sim 0.02$\,pc ($\sim 10^{\prime\prime}$). The observational center of the cloud is positioned at RA(J2000)\,=\,14h\,48m\,29.00s, Dec(J2000)\,=\,--65$^{\circ}$\,15$^{\prime}$\,54.0$^{\prime \prime}$, with a projected distance of $\sim 0.63$\,pc from HD\,130079. Spectral regions range from a distance of $\sim$\,0.03 to $\sim$\,0.18\,pc from HD\,130079. }
    \label{Regions}
\end{figure}

Yet, the emission produced in the surroundings of young stars, is not the only agent capable of ionizing the ISM or of causing grain destruction. In fact, lower-energy cosmic rays are often considered as an equally relevant source of ionization, in particular in the context of dense molecular clouds and the Galactic Center region \citep[e.g.,][]{Prasad83,Padovani09,Caselli12, Indriolo13,Indriolo15}. In this context: large PAHs can substantially modify the ionization balance in their environment via a charge transfer from interstellar ions (leading to the creation of PAH cations), or electron attachment (producing PAH anions), and subsequent mutual neutralization \citep{Bakes98,Dalgarno06,Verstraete11}. In addition cosmic rays should be also considered as a relevant agent responsible for destroying PAHs, especially the smaller ones, and especially in dense clouds, where the competing processes of PAH destruction by interstellar shocks or UV emission, are rather inefficient \citep{Micelotta11}.

In this paper, we discuss the PAH features of a region in at the eastern edge of DC\,314.8--5.1, mapped with the Spitzer Space Telescope \citep{Werner04}. Our work is motivated by the previous study by \citet{Whittet07}, who identified the system as a compact and dense ``dark globule'', at the onset of low-mass star formation \citep[see in the context the review by][]{Bergin07}. The distance and the parameters that depend on distance, could be constrained relatively precisely by the serendipitous association of the cloud with a B-type field star, which illuminates a reflection nebula near its eastern boundary. The Spitzer Space Telescope observations analyzed here, provide us the opportunity to also study the PAH features, at varying distance from the star, and at the same time from the center of the globule. In this way, we seek to diagnose the ionization level conditions and the dust grain distribution within the dense molecular cloud, \emph{before} any formation of stars, and so not affected by the presence of young stellar objects.

The paper is organized as follows. In the next Section\,\ref{sec:overview}, we summarize the established properties of the DC\,314.8--5.1. In the following Section\,\ref{sec:obs}, we introduce the Spitzer Space Telescope observations targeting the system, and next in Section\,\ref{sec:data} we discuss the analysis performed on the Spitzer data. In Section\,\ref{sec:results} we present the analysis results regarding the PAH features and ratios; these results are further discussed in Section\,\ref{sec:discussion} and concluded in the final Section\,\ref{sec:conclusions}.

\section{System Overview}
\label{sec:overview}

DC\,314.8--5.1 is a dark globule $\sim 8^{\prime} \times 5^{\prime}$ in angular size, located approximately 5\,degrees below the Galactic Plane, RA(J2000)\,=\,14h\,48m\,29.00s, Dec(J2000)\,=\,--65$^{\circ}$\,15$^{\prime}$\,54.0$^{\prime \prime}$, in the constellation of Circinus; see Figure\,\ref{DSS}. The system was originally identified in the \citet{Hartley86} catalogue of southern dark clouds and was selected from that catalogue for further study. The globule is of particular interest due to the association of field star, HD\,130079, embedded in its eastern boundary. HD\,130079 is a $10^{th}$ magnitude, normal main sequence star, with a well-established spectral type of B9V \citep[][and references therein]{Whittet07}; it is illuminating a reflection nebula on the eastern boundary of the cloud.

The association of the star HD\,130079 with DC\,314.8--5.1 allows important physical characteristics of the system to be estimated. A review of the physical properties of HD\,130079 and the dust responsible for extinction in the line of sight was carried out by \citet{Whittet07}, with the goal of better constraining the distance to the star (and hence, to the cloud). Since this work was completed, new and far more accurate constraints on the distance to HD\,130079 have become available as the result of parallax measurements by the Gaia mission. The parallax value for the star in the EDR3 data release \citep{Gaia21} is $2.2981 \pm 0.0194$\, mas. This result corresponds to a distance of $435\pm 4$\,pc, placing it at a significantly greater distance compared with the photometric estimate of $342 \pm 50$\,pc deduced by \citet{Whittet07}. The most likely cause of the discrepancy is an over-estimate of the ratio of total to selective extinction in the earlier work, possibly resulting from contamination of the photometry by a fainter, redder star in the aperture. With the revised distance, the angular dimensions of the dark cloud ($\sim 8^{\prime} \times 5^{\prime}$; Figure\,\ref{DSS}) correspond to a projected linear size of $\sim 0.9$\,pc\,$\times 0.6$\,pc. Using this result and other observed properties of the dark cloud, the mean core number density of hydrogen gas is $\sim 10^4$\,cm$^{-3}$, and the total mass of the cloud is $\sim 160 M_{\odot}$ \citep[see][]{Hetem88,Whittet07}.

DC\,314.8--5.1 appears to be in a quiescent state prior to the start of star formation. There has been significant interest in verifying the validity of these so called ``starless cores'', but according to a study performed by \citet{Kirk07} the probability of a misidentified core (i.e. that a ``starless core'' is not in fact starless but host to some stellar object) is low ($\sim 5\%$). Proper verification for a starless core can be performed by utilizing the Multiband Imaging Photometer (MIPS) onboard the Spitzer Space Telescope, to search for sources embedded in a system \citep{Kirk07}. A stellar consensus utilizing the Two Micron All-Sky Survey \citep[2MASS;][]{2MASS} yielded only two potential Young Stellar Object (YSO) candidates in DC\,314.8--5.1, one of which was excluded as a old star with significant dust reddening, out of a sample of 387 sources \citep{Whittet07}. The globule, therefore, was determined \emph{not} to be a site of vigorous star formation, and as such is a valuable system to study the dynamics of a pre-stellar globule.

\section{Spitzer Observations}
\label{sec:obs}
 
Observational data for this work were provided by the Spitzer Space Telescope (Proposal ID\,50039; P.I.: D.~Whittet), acquired by the InfraRed Spectrograph \citep[IRS;][]{Houck04} Mapping mode of the reflection nebula near the field star HD\,130079. The IRS mapping observations, obtained through the Spitzer Heritage Archive, are used to investigate the relationship between the radiation from the field star and the corresponding PAH features within the globule.

IRS mapping observation of the reflection nebula was obtained in both the short-low (SL) and long-low (LL) modules of IRS and covers the wavelength range of 5.13 to 39.0\,$\mu$m. These data have a resolving power of $\simeq$\,60\,--127 and a signal-to-noise ratio S/N\,$\simeq 20$. We achieve the high S/N in a relatively short amount of time: $2\times 60$\,s in SL and $2 \times 14$\,s in LL. The slits are overlapped by $1/2$ of the slit dimensions in both directions which mitigates artifacts due to bad and rogue pixels and results in a reduction of errors in flat-fielding.

\section{Data Analysis} 
\label{sec:data}

\subsection{Reduction of Data Products}

\begin{figure}[h!]
    \centering
    \includegraphics[width=0.9\columnwidth]{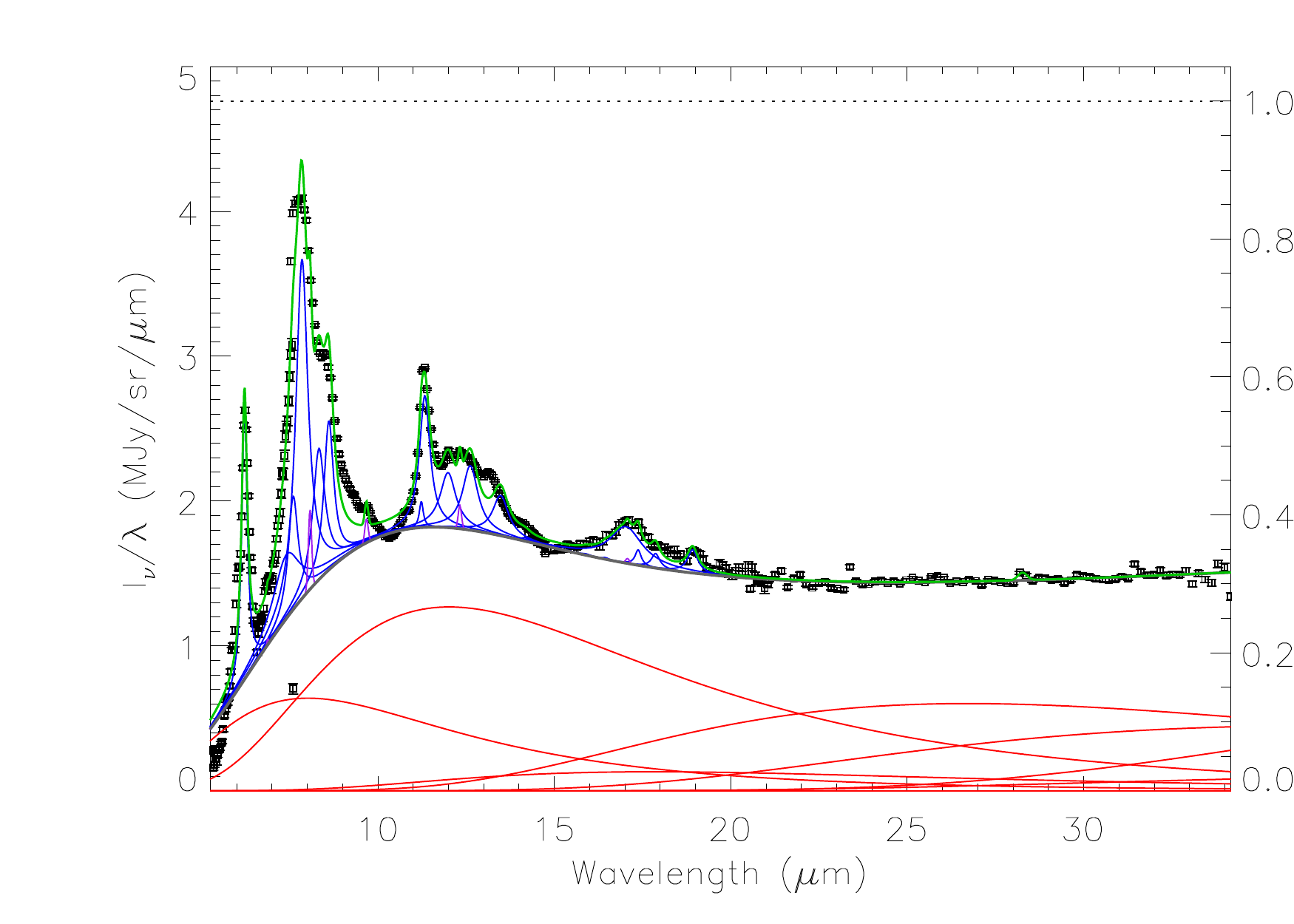}
    \includegraphics[width=0.9\columnwidth]{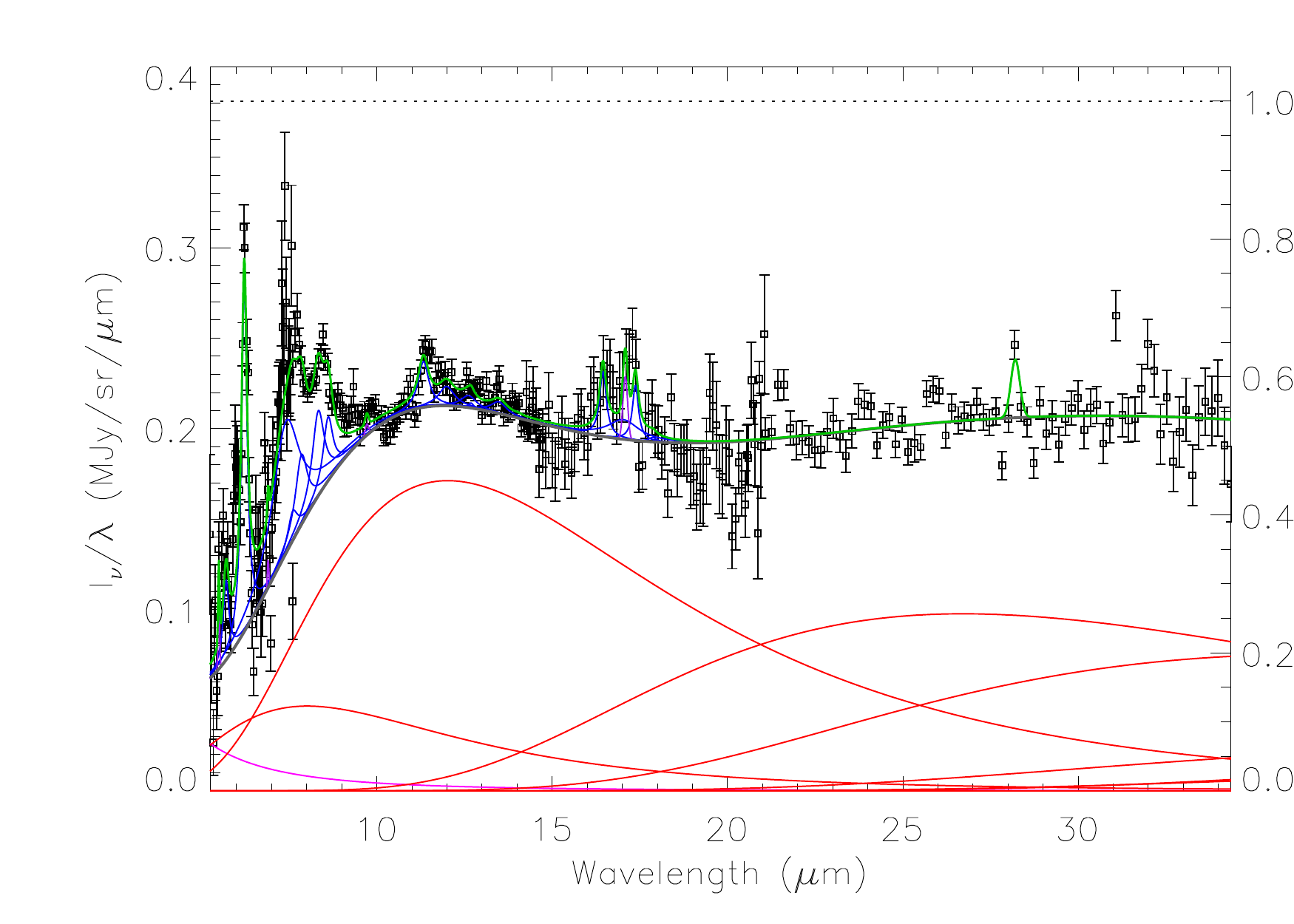}
    \includegraphics[width=0.9\columnwidth]{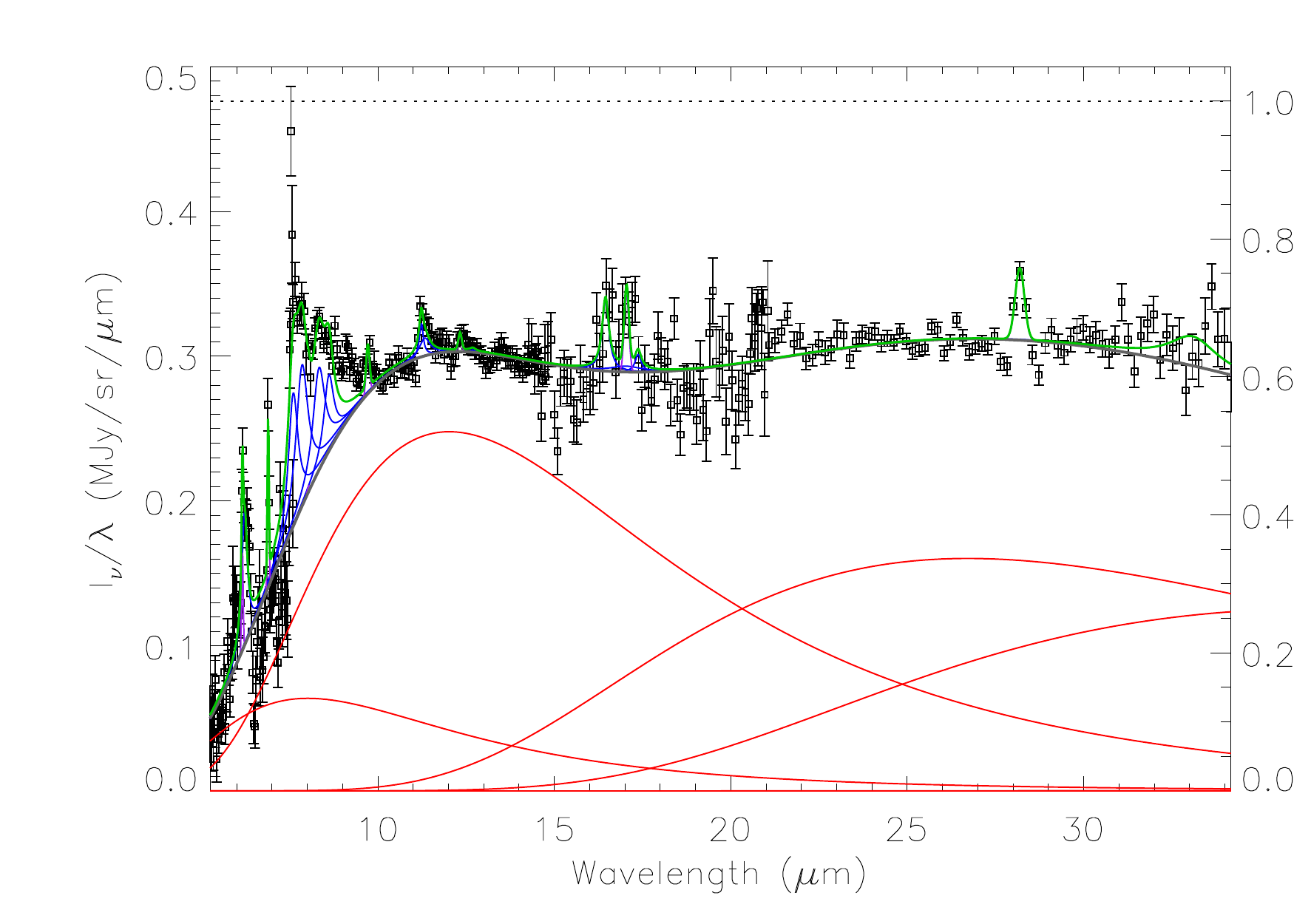}
    \caption{Fully fitted spectra extracted from the three regions at the projected distances of 0.03\,pc, 0.09\,pc, and 0.16\,pc away from the illuminating field star, HD\,130079 (upper, middle, and lower panels, respectively),  see Figure\,\ref{Regions} for the physical extent of the extraction regions for each spectra. Spectral fitting of the regions was performed with PAHFIT \citep{PAHFIT}. The green curve represents the best fit, red curves represent the thermal dust continuum components, thick gray curves indicate the total dust and stellar continuum, blue curves indicate the fitted PAH features, violet curves indicate atomic and molecular spectral lines, and green curves represent the fully fitted model. The dotted black curves (present in the middle and bottom panels) denotes the level of extinction (fully mixed) present (relative fraction presented on the right vertical axis).}
    \label{Lowres}
\end{figure}

The data reduction procedure for the SL and LL spectroscopic data followed was similar to the CUBISM recipe outlined in the Spitzer Data Cookbook\footnote{\url{https://irsa.ipac.caltech.edu/data/SPITZER/docs/}}, Recipe 10, with parameters appropriate for our dataset. The CUBISM standard bad pixel generation was applied with the following values: ${\sigma}_{trim} = 7$, minimum bad-fraction equal to $0.5$ for global bad pixels and equal to $ 0.75$ for record bad pixels. Due to the source being a low luminosity object, additional bad pixels were flagged visually before spectral extraction. 117 extraction regions were defined of dimensions $2 \times 2$ pixels around the field star, HD\,130079, within the reflection nebula, as visualized in Figure\,\ref{Regions}. Regions were selected such that the center of each new region was the edge of the previous region to account for subtle changes over distance. Spectral extraction was performed by selecting a region beginning on the star, of four pixels, and shifting the extraction region one pixels for each new extraction (in both vertical and horizontal directions). Projected distances from the field star were calculated from the center of each selection cube. Each cube (i.e., each array of four neighbouring pixels, see Figure\,\ref{Regions}), spans an angular width of $\sim 10^{\prime\prime}$, which at a distance of $435 \pm 4$\,pc (a systematic error of less than 1\%) translates to a spatial size of $\sim$\,0.021\,pc. In order to extract spectra for the entire wavelength range, four cubes were built for each spatial extraction region, one for each spectral mode (i.e. SL1, SL2, LL1, LL2) \citep{Cubism}.

\begin{figure*}[ht]
   \centering
   \includegraphics[width=0.75\textwidth]{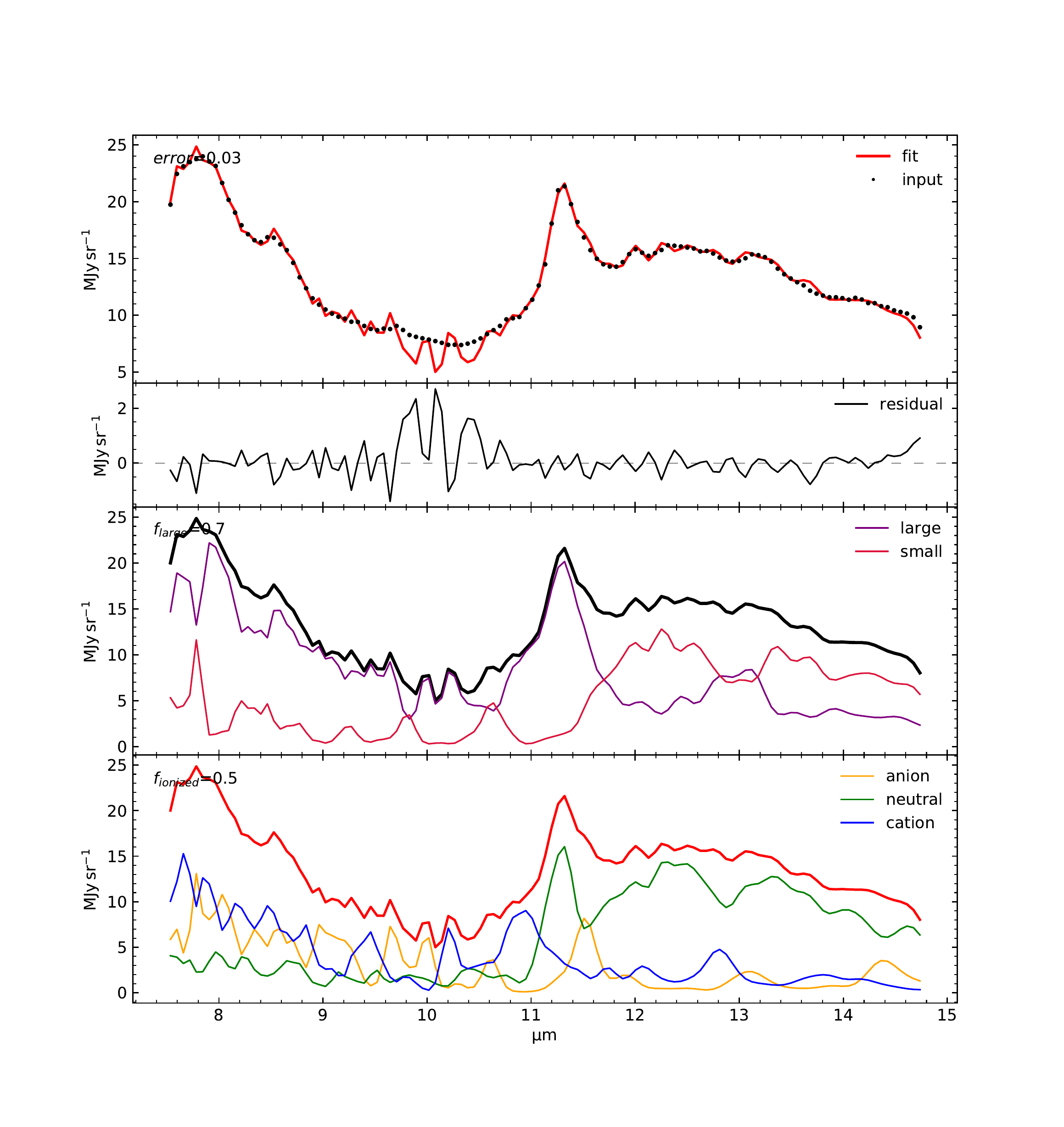}
  \vspace{-1cm}
   \caption{Pypahdb \citep{PYPAHDB} fit of the SL1 mode from the nearest spectral region to HD\,130079 at a projected distance of 0.03\,pc. From top to bottom the individual panels represent: the fully fitted spectrum  with data points marked as black dots with respective errors, note here that error bars are significantly small as to not be easily visible}, residuals in the final fit, the breakdown of the PAH species sizes into large ($N_{\rm C}>40$) and small species, and the breakdown of the cation, neutral, and anion species.
         \label{pahdb}
   \end{figure*}

The combination of spectral modes (i.e. SL1, SL2, LL1, and LL2) was performed utilizing python coding routines, developed with Python (3.5.9 distribution\footnote{\url{https://docs.python.org/3/}}), and standard averaging methods involving propagation of errors. Differences in the continuum of the SL and LL modes (prevalent in low-intensity regions) were taken into account by scaling the LL modes to the SL modes.

\begin{figure*}[ht]
   \centering
   \includegraphics[width=\textwidth]{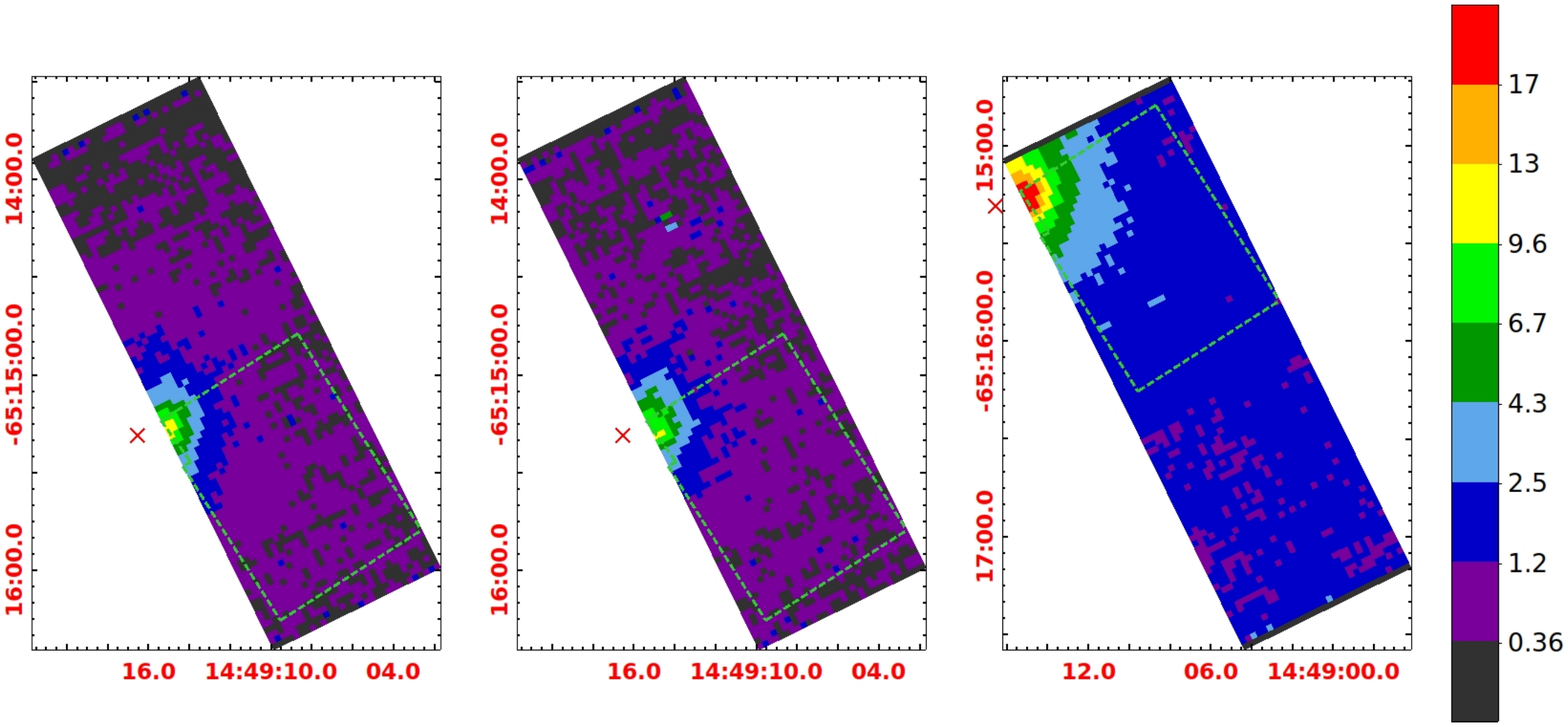}
   \vspace{-3cm}
   \caption{Square root scaled intensity maps of the continuum-subtracted PAH features: 6.2 (left), 7.7 (middle), and 11.3\,$\mu$m (right).  The colorbar is set with a range across all the features from 0.1-21.5, with the maximum set by the 11.3\,$\mu$m feature and the minimum intensity set by the 6.2\,$\mu$m feature map. In all images orientation is the same as in Figure\,\ref{DSS}, North is orientated up and East is orientated to the left. The green dashed lines delineate the extraction regions with the red cross indicating the optical center of HD\,130079.}
         \label{Intensity}
   \end{figure*}
   
\subsection{Fitting Process for Data Products}

Spectral decomposition of the IRS Map observations was performed with the PAHFIT spectral fitting software \citep{PAHFIT}. PAHFIT utilizes a model consisting of starlight, thermal dust continuum in fixed temperature bins, resolved dust features, prominent emission lines, and silicate dust extinction. This fitting was performed on all 117 extracted low resolution spectra taken at varying distances away from HD\,130079 (and therefore increasing distance to the cloud center; see Figure\,\ref{Regions}). The PAHFIT process results in images for each fit as well as text files denoting the central wavelength, intensity, and the respective errors for each line fit, and also levels for stellar and dust continuum fits. Specifically PAHFIT measures the Starlight, Dust Continuum at 300, 200, 135, 90, 65, 50, 40, and 35 degrees Kelvin, 18 emission lines (H$_2$, Ar, S, Fe, Si, and Ne), 25 Drude profiles to fit the PAH features, and the extinction level. Due to the high excitation for the atomic lines, we do not expect/nor do we see any evidence of their presence in our sample and so all atomic lines are removed by fixing their normalization to zero during the fitting procedure \citep[see][]{Herbst09}.

Figure\,\ref{Lowres} shows for illustration the three spectra extracted at three various distances from the star, after PAHFIT fitting. Data points are presented by black squares (with respective error bars), red curves indicate the stellar continuum, thick gray curves indicate the dust and stellar continuum, blue curves indicate the fitted dust features, violet curves indicate atomic and molecular spectral lines, the dotted curves indicate components diminished by fully mixed extinction, and the green curves represent the fully fitted model \citep{PAHFIT}. As is evident in Figure\,\ref{Lowres}, the $5-20$\,$\mu$m range is dominated by emission features, whereas the spectrum is almost featureless at longer wavelengths; we therefore focus on  the $5-20$\,$\mu$m spectrum in the following sections. Specifically we note the strong ionized $6.2$ and $7.7$\,$\mu$m features, as well as the ionized $8.6$\,$\mu$m feature, in addition to a strong $11.2-11.3$\,$\mu$m neutral PAH feature. 

Complementary fitting of the extracted spectra were performed utilizing the pypahdb\footnote{\url{ https://pahdb.github.io/pypahdb/}} package \citep{PYPAHDB}. Unlike the PAHFIT routine, which decomposed the 5-35\,$\mu$m low-resolution Spitzer spectra into the dust continuum emission (modeled as a superposition of modified blackbodies), PAH features (fitted with phenomenological Drude profiles), and atomic/molecular lines (approximated as Gaussians), the pypahdb package utilizes the NASA Ames PAH IR Spectroscopic Database \citep[see][for an updated summary]{PAHDB}, to directly extract the ionization fraction and the size breakdown for the PAH molecules within the analyzed region by means of fitting the observed spectrum against the library of computed PAH spectra which contains data on thousands of PAH species.

\begin{deluxetable}{crr}[th!]
\tabletypesize{\footnotesize}
\tablecaption{PAH Drude Profiles}
\tablewidth{0pt}
\tablehead{
\colhead{Feature} & \colhead{Wavelength }& \colhead{Max Intensity}  \\
\colhead{~~} & \colhead{[$\mu$m]}& \colhead{[MJy\,sr$^{-1}$]}  }
\startdata
\hline
  \multicolumn{3}{c}{{ \textbf{The ``Inverse Power-law'' Category}}}\\
PAH $6.2$ & $6.22$ & $12.3\pm0.002$\\
PAH $7.4$ & $7.42$ & $4.26\pm0.009$\\
PAH $11.3$ & $11.3$ & $10.2\pm0.003$ \\
\hline
  \multicolumn{3}{c}{{ \textbf{The ``Inverse Power-law + Plateau'' Category}}}\\
PAH $7.6$ & $7.6$ & $5.39\pm0.010$ \\
PAH $7.8$ & $7.85$ & $18.5\pm0.002$  \\
PAH $8.3$ & $8.33$ & $7.69\pm0.003$  \\
PAH $8.6$ & $8.61$ & $8.84\pm0.003$\\
PAH $11.2$ & $11.2$ & $2.05\pm0.023$\\
PAH $17.0$ & $17.0$ & $4.67\pm0.020$\\
PAH $17.9$ & $17.8$ & $2.06\pm0.046$\\ 
PAH $18.9$ & $18.9$ & $3.46\pm0.028$\\
\enddata
\tablenotetext{}{{\bf Col(1)} --- PAH Drude profile designation; {\bf Col(2)} --- Central wavelength as fit by PAHFIT; {\bf Col(3)} --- Maximum intensity out of the 117 regions as fit by PAHFIT.}{}
\label{table:pahtrend}
\end{deluxetable}

For our study, we utilized pypahdb for each of our sampled regions for the Spitzer SL1 mode (i.e. 7.5--15\,$\mu$m).  We do not include the SL2 mode in this fitting as the inclusion of the additional segment from SL2 (6.5--7.5\,$\mu$m) greatly reduces the quality of the fit, increasing residuals to $>10$\,MJysr$^{-1}$. Figure\,\ref{pahdb}, presents the pypahdb fit for the extraction region located nearest to the field star, HD\,130079, at a distance of 0.03\,pc, showing the respective breakdown of ionization and size, where large species are defined as $N_{\rm C}>40$. The resulting spectral fits have a high SNR\,$>5$, with only five regions having $5>$\,SNR\,$>3$, and only one region had an average residual equating a SN lower than 3\,$\sigma$ and as such was removed from any further analysis.

\subsection{Extraction of Intensity Maps}

Intensity maps were extracted from the Spitzer spectral cubes with Cubism for selected features \citep{Cubism}. For each map the continuum is defined as the regions on either side of the feature, and removed, taking particular care in this procedure to not include any neighbouring features or lines. The peak of the desired feature is then selected. This procedure produces a continuum-subtracted map of the averaged surface brightness in units of MJy\,sr$^{-1}$. In Figure\,\ref{Intensity} we present the resulting 6.2, 7.7 and 11.3\,$\mu$m PAH feature maps, created by defining the feature with ranges for each peak as follows:  6.0--6.5\,$\mu$m, 7.3--7.9\,$\mu$m, 11.2--11.4\,$\mu$m.
   
\begin{deluxetable}{crr}[th!]
\tabletypesize{\footnotesize}
\tablecaption{Integrated PAH Features}
\tablewidth{0pt}
\tablehead{
\colhead{Feature} & \colhead{Wavelength }& \colhead{Max Power} \\
\colhead{~~} & \colhead{[$\mu$m]}& \colhead{[Wm$^{-2}$\,sr$^{-1}$ $\times10^{-9}$]} }
\startdata
\hline
  \multicolumn{3}{c}{{ \textbf{The ``Inverse Power-law'' Category}}}\\
PAH $6.2$ & $6.2-6.3$ & $279\pm1.38$\\
PAH $11.3$ & $11.2-11.4$ & $146\pm0.712$ \\
\hline
  \multicolumn{3}{c}{{ \textbf{The ``Inverse Power-law + Plateau'' Category}}}\\
PAH $7.7$ & $7.3-7.9$ & $1070\pm5.41$  \\
PAH $8.6$ & $8.6-8.7$ & $189\pm1.35$\\
PAH $17.0$ & $16.4-17.9$ & $99.5\pm2.87$\\
\enddata
\tablenotetext{}{{\bf Col(1)} --- PAH feature designation; {\bf Col(2)} --- Central wavelength as fit by PAHFIT; {\bf Col(3)} --- Maximum power measured out of 117 regions as fit by PAHFIT.}{}
\label{table:mergetrend}
\end{deluxetable}

\subsection{Extinction Profile}

For the inspection of line features within the dark globule, DC\,314.8--5.1, it is important to comment on the extinction within the sampled region. Given in \citet{Whittet07} the lower limit of the extinction of the cloud core is $A_V\gtrsim 8.5$\,mag, however, this is not the region that is sampled in our study. We are investigating the outskirts of the cloud from 0.03--0.18\,pc projected distance from HD\,130079, which corresponds to a distance range from the core of DC\,314.8--5.1 of 0.3--0.45\,pc. Given however the extinction of 8.5 in the core of the cloud, the established relation $N_{H_2}/A_V \simeq 9.4\times 10^{20}$\,cm$^{-2}$\,mag$^{-1}$ \citep{Hetem88}, and assuming the standard mass density profile for the cloud $\rho (r) \propto r^{-1.3}$ \citep{Cernicharo85}, we can estimate the extinction level within a certain distance probed in this paper, namely 0.3--0.45\,pc from the center of the cloud, as 1.0--1.8\,mag (corresponding to the $H_2$ column density of $\sim 1.0-1.7 \times 10^{21}$\,cm$^{-2}$). The extinction within the sampled distances can therefore affect the PAH emission of the cloud, although we should still be able to probe the full depth of the targeted segment of the system. Obviously, given the simplicity of the above-mentioned estimate, it is possible that the extinction in this region may be higher and we may only be seeing the surface of the cloud contributing to the PAH emission seen.

\begin{figure*}[!th]
\centering
	\includegraphics[width=0.32\textwidth]{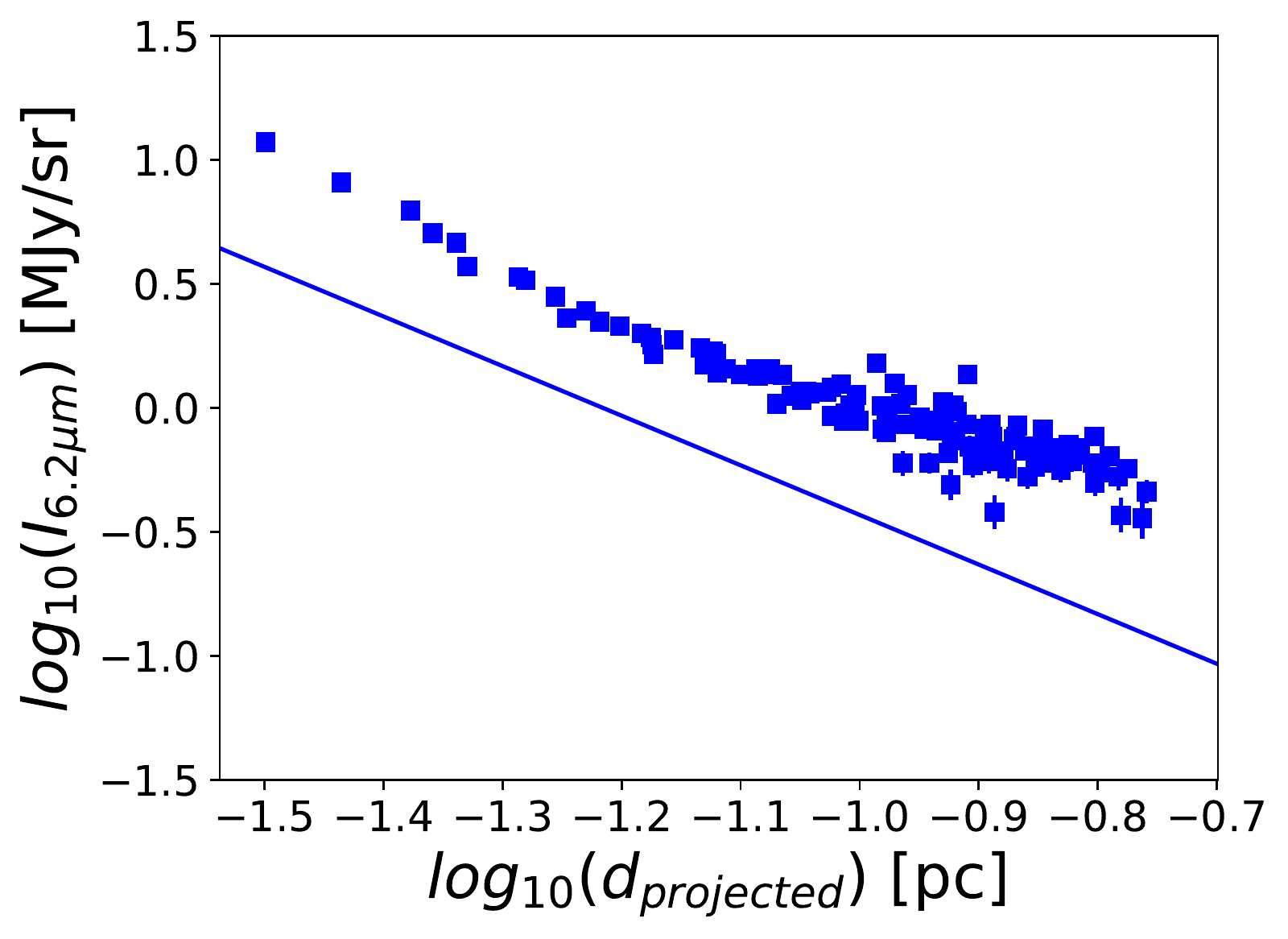} 
	\includegraphics[width=0.32\textwidth]{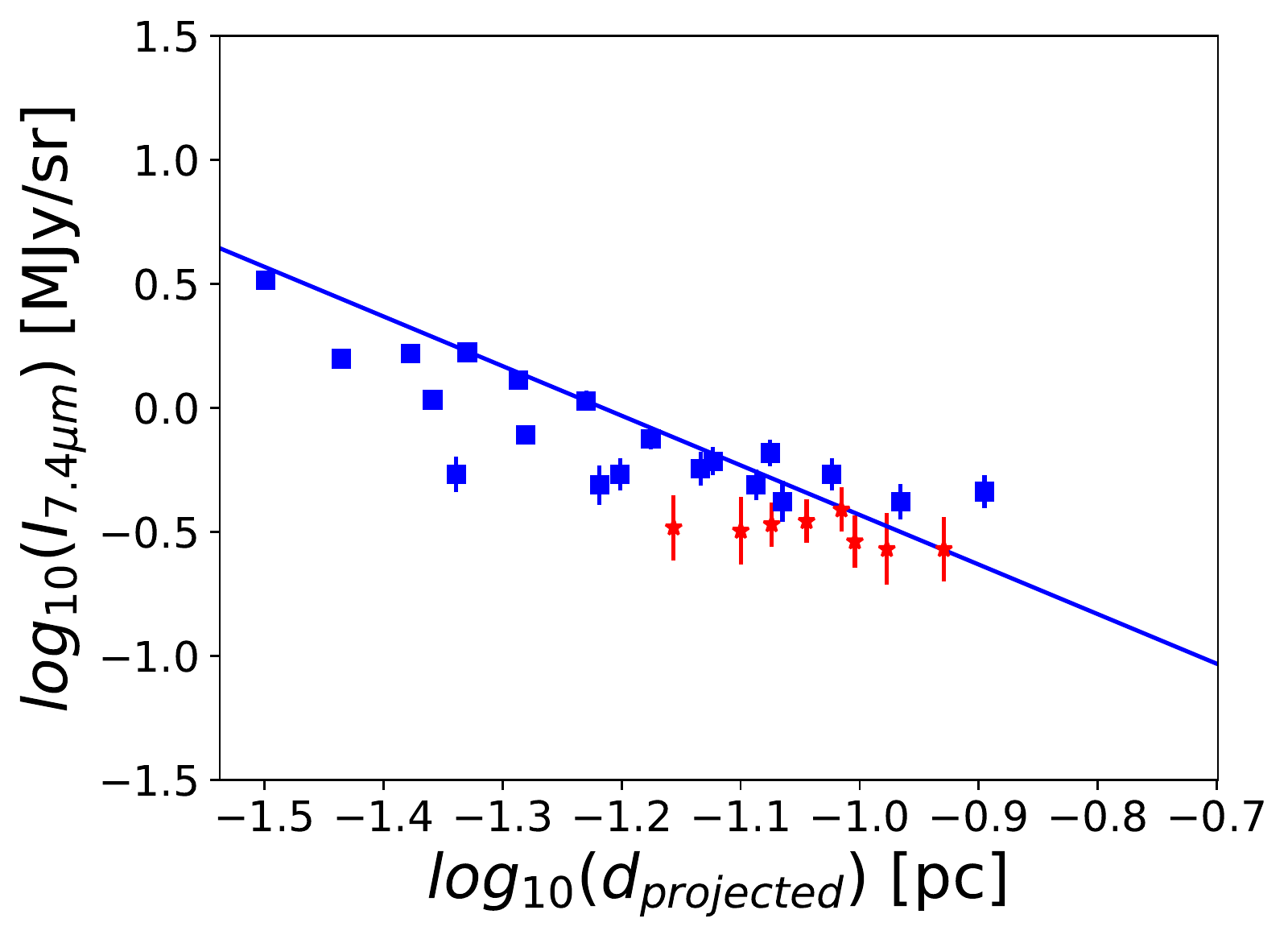} 
	\includegraphics[width=0.32\textwidth]{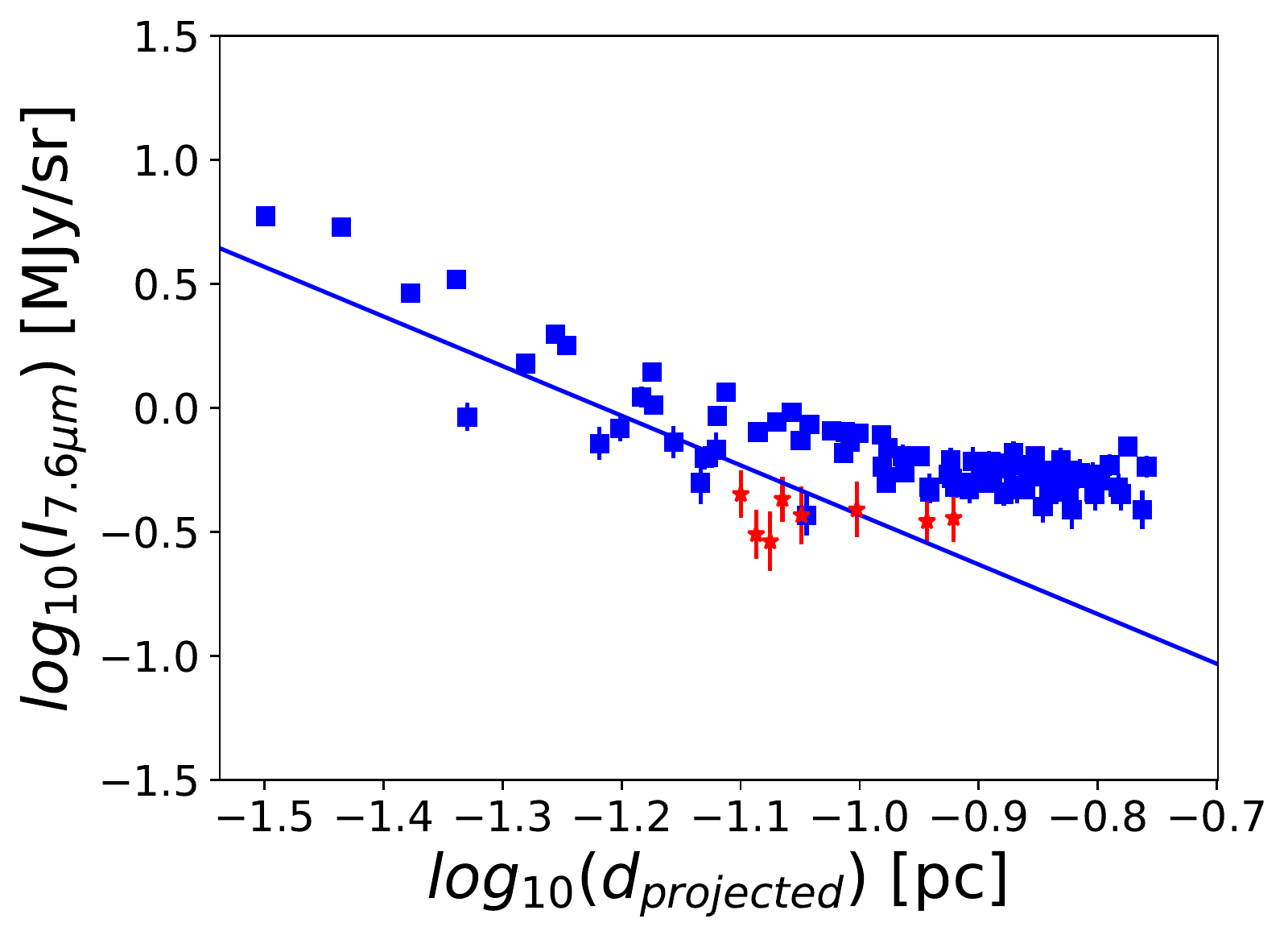} 
	\includegraphics[width=0.32\textwidth]{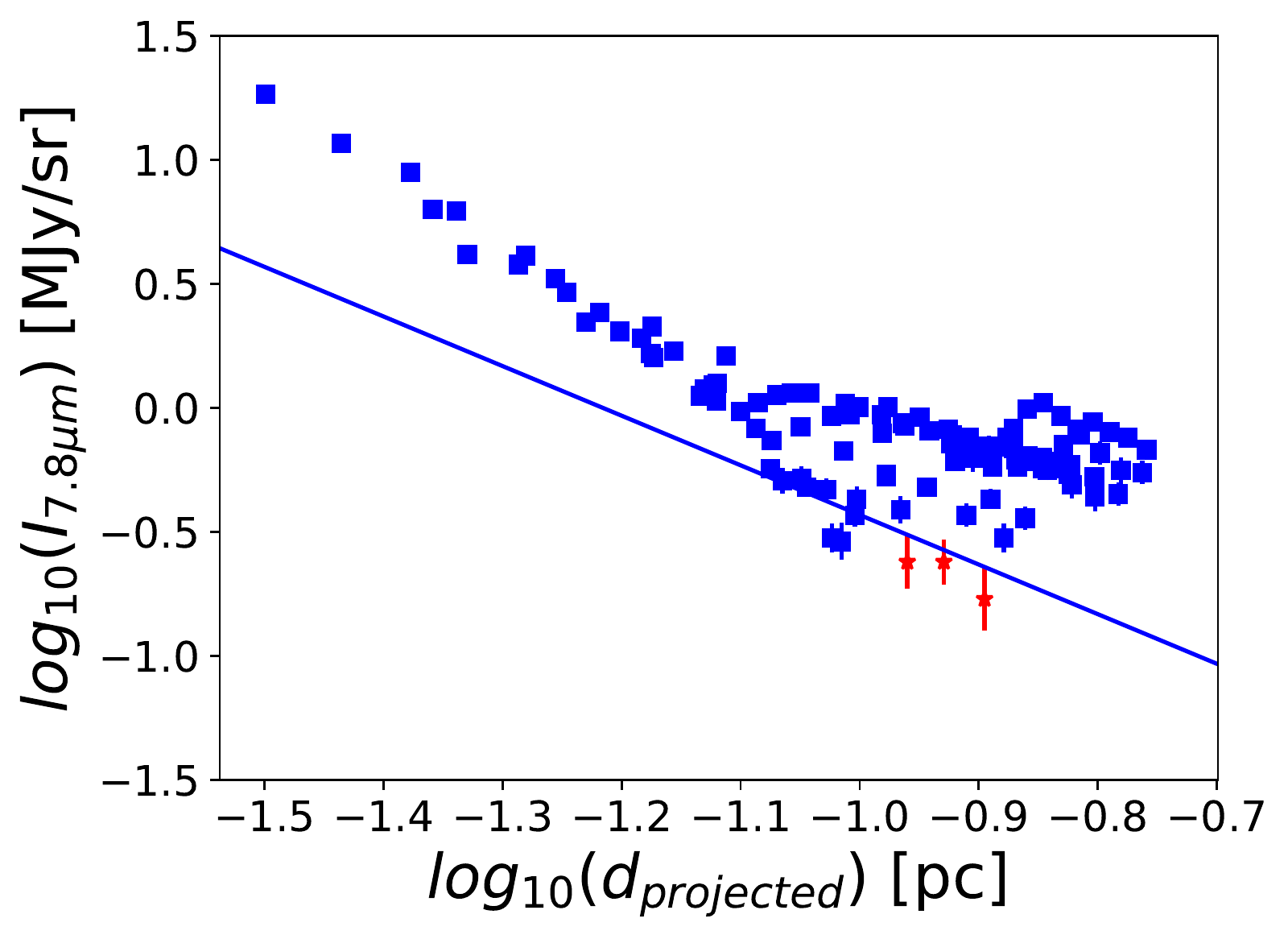} 
	\includegraphics[width=0.32\textwidth]{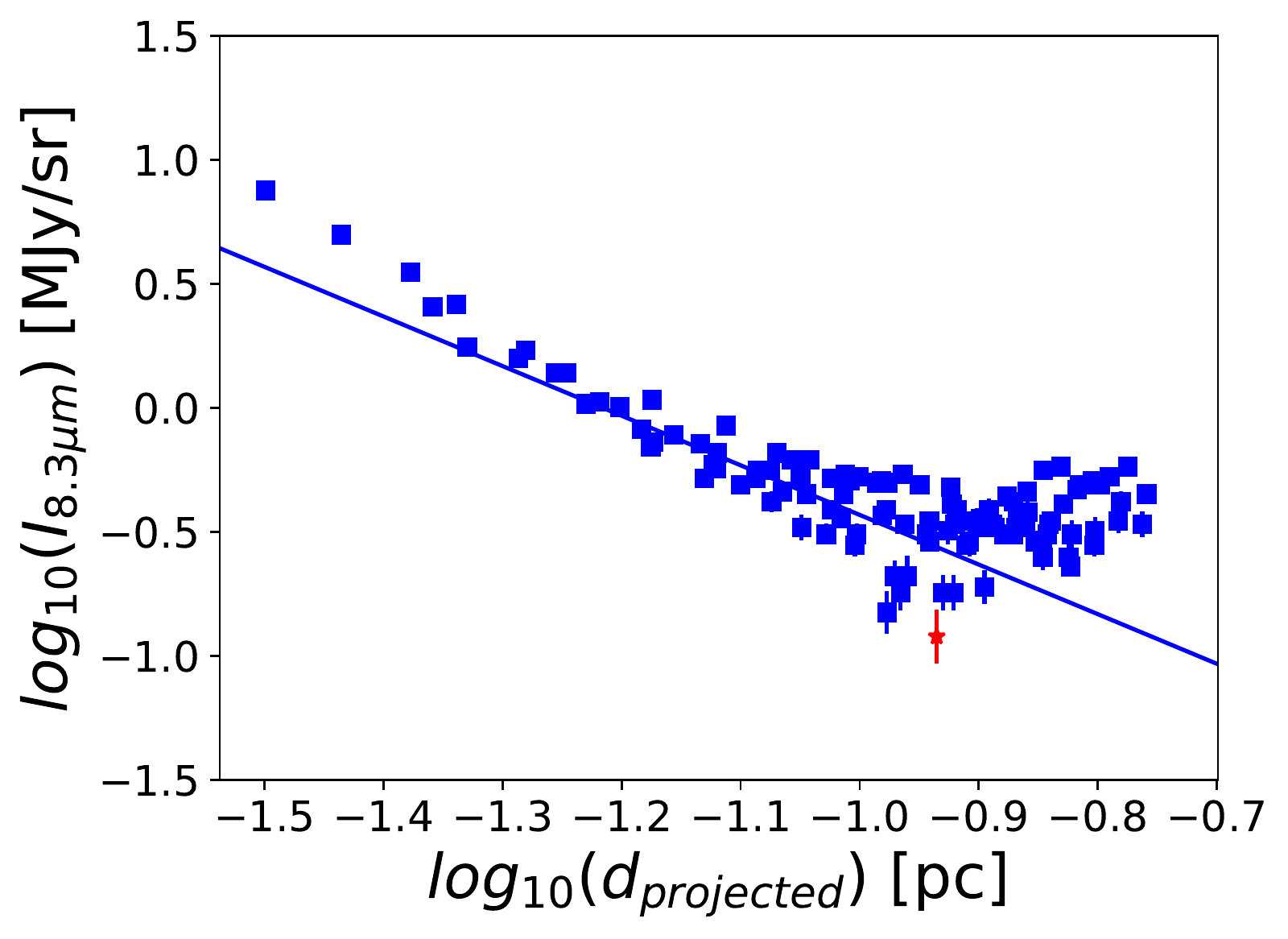} 
	\includegraphics[width=0.32\textwidth]{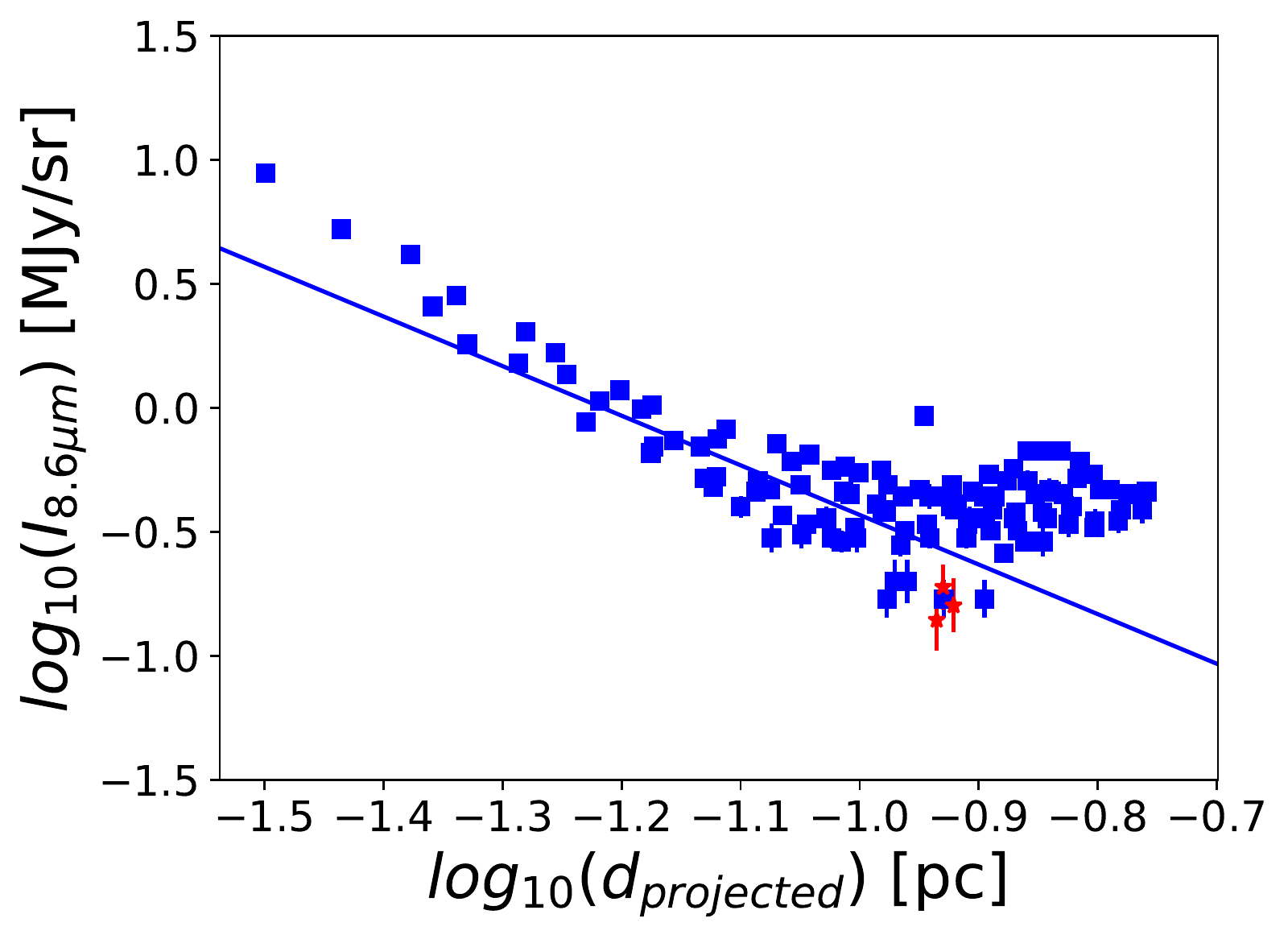} 
	\includegraphics[width=0.32\textwidth]{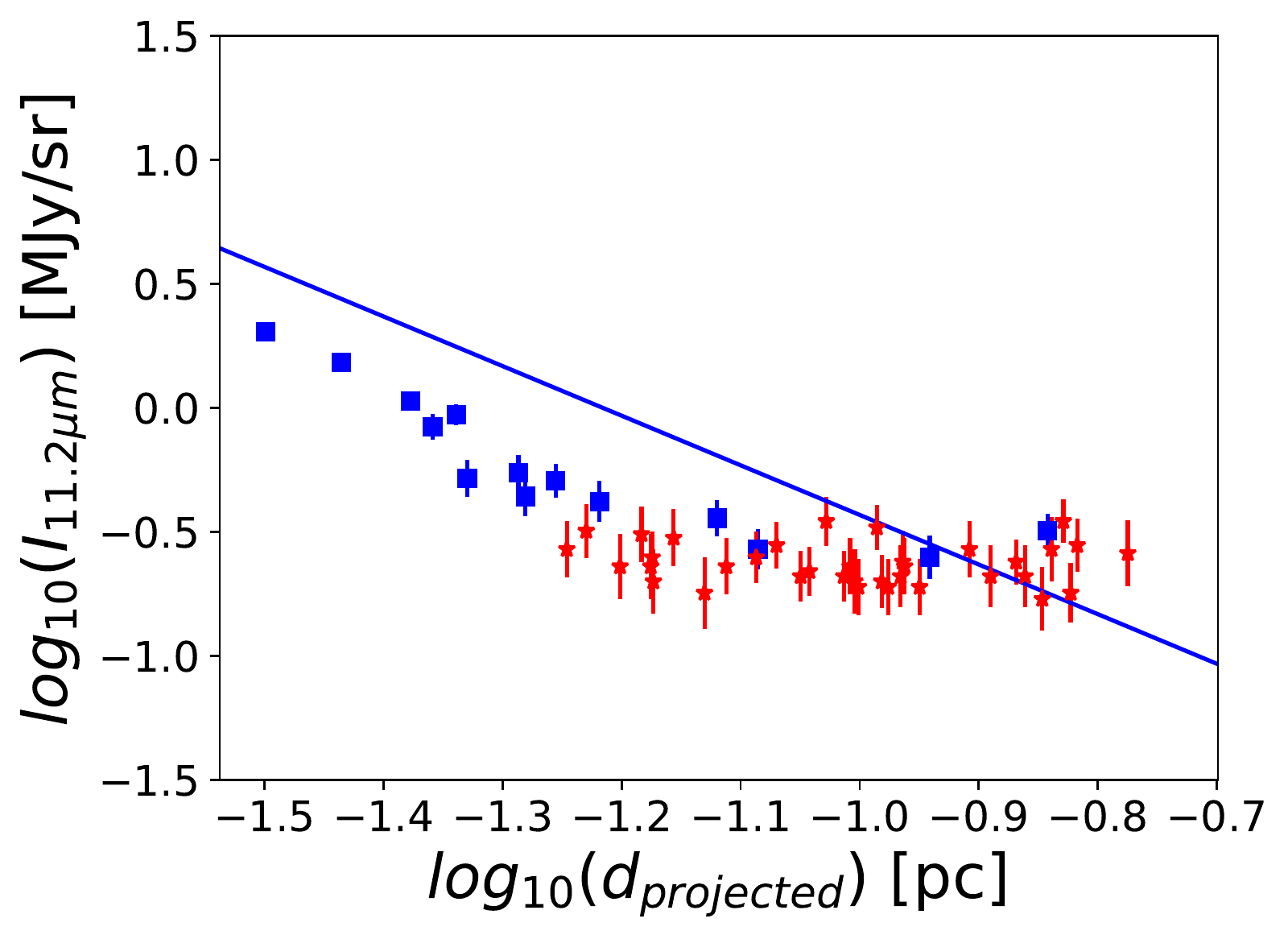} 
	\includegraphics[width=0.32\textwidth]{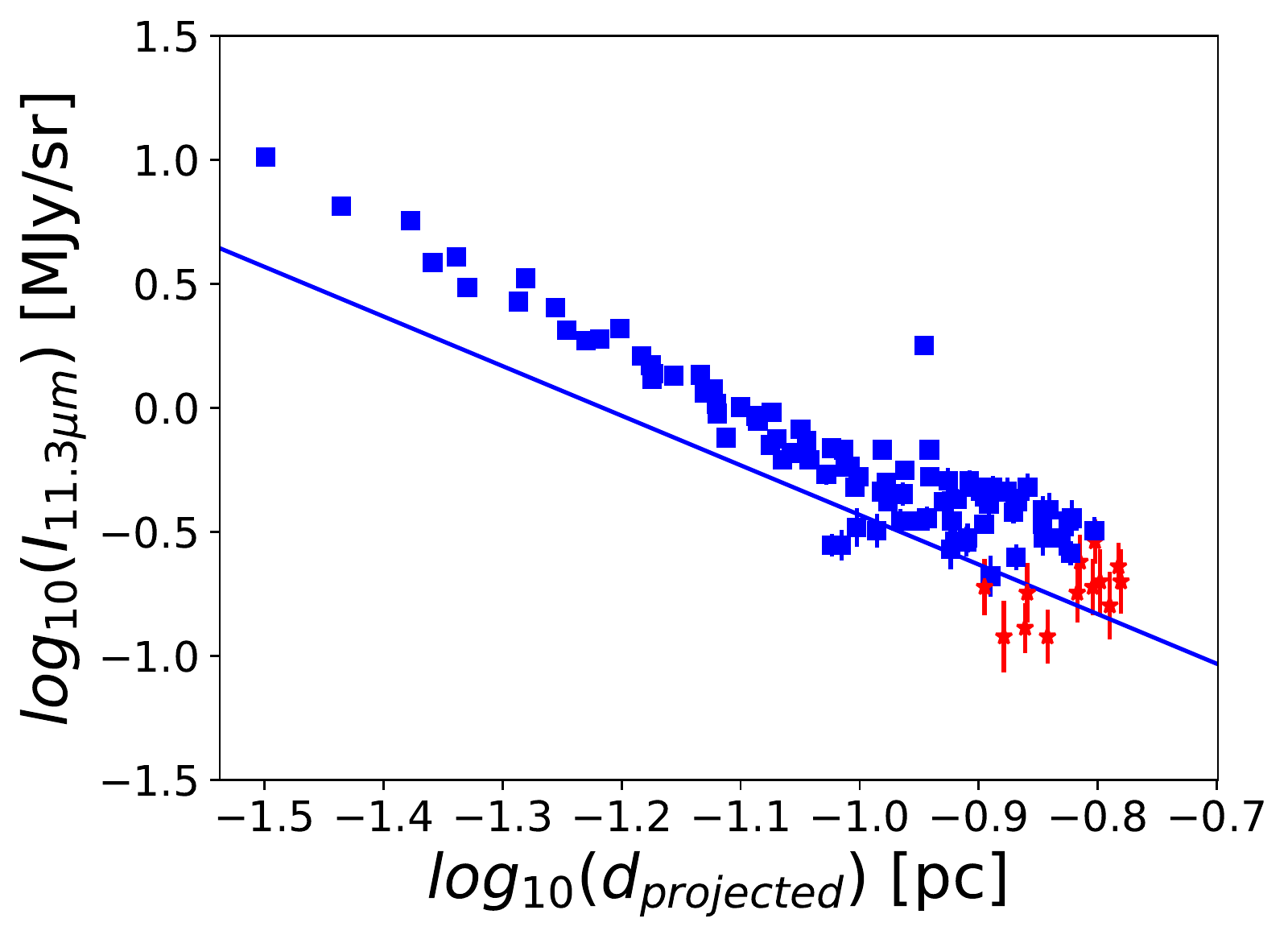} 
	\includegraphics[width=0.32\textwidth]{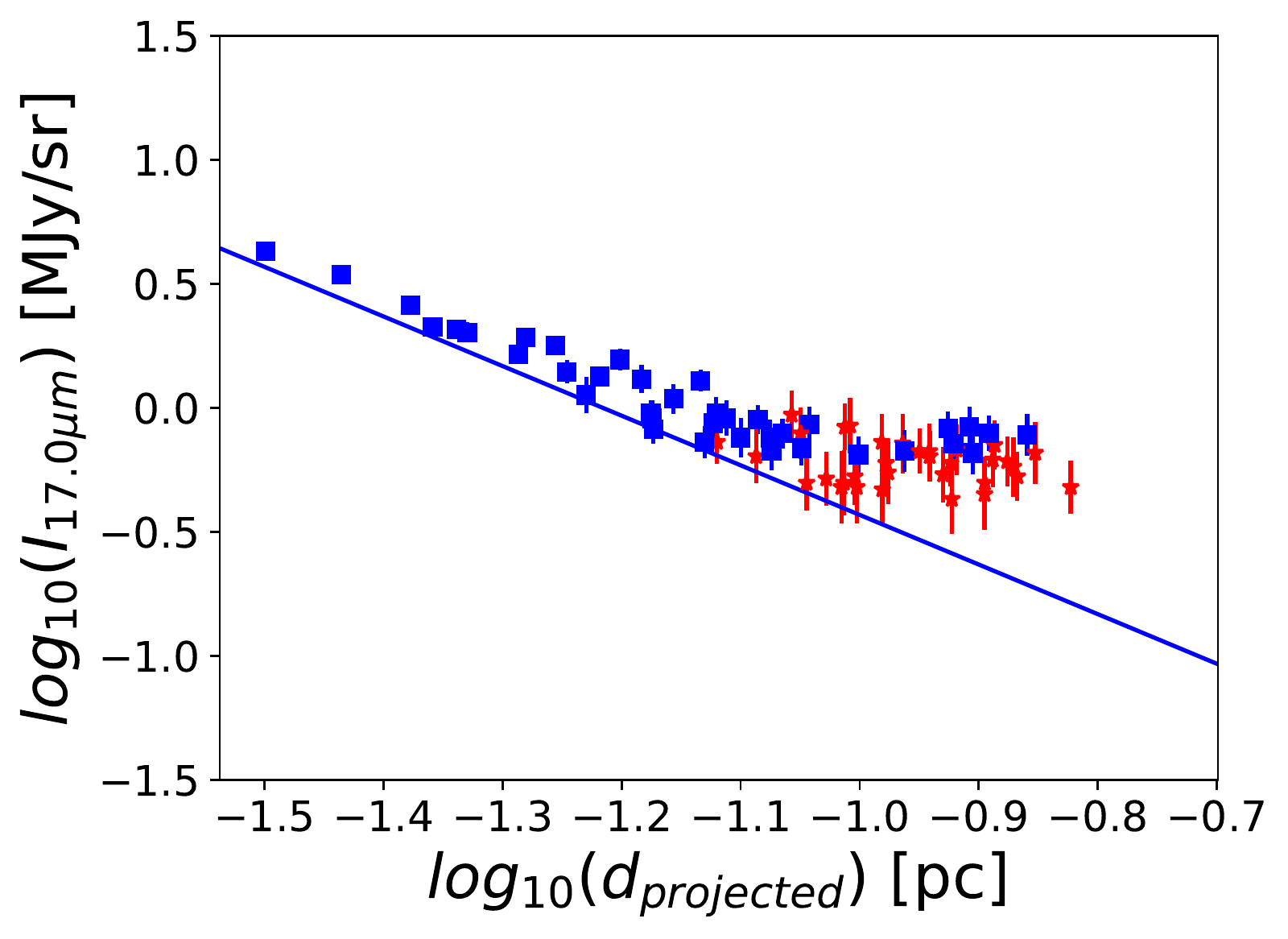} 
	\includegraphics[width=0.32\textwidth]{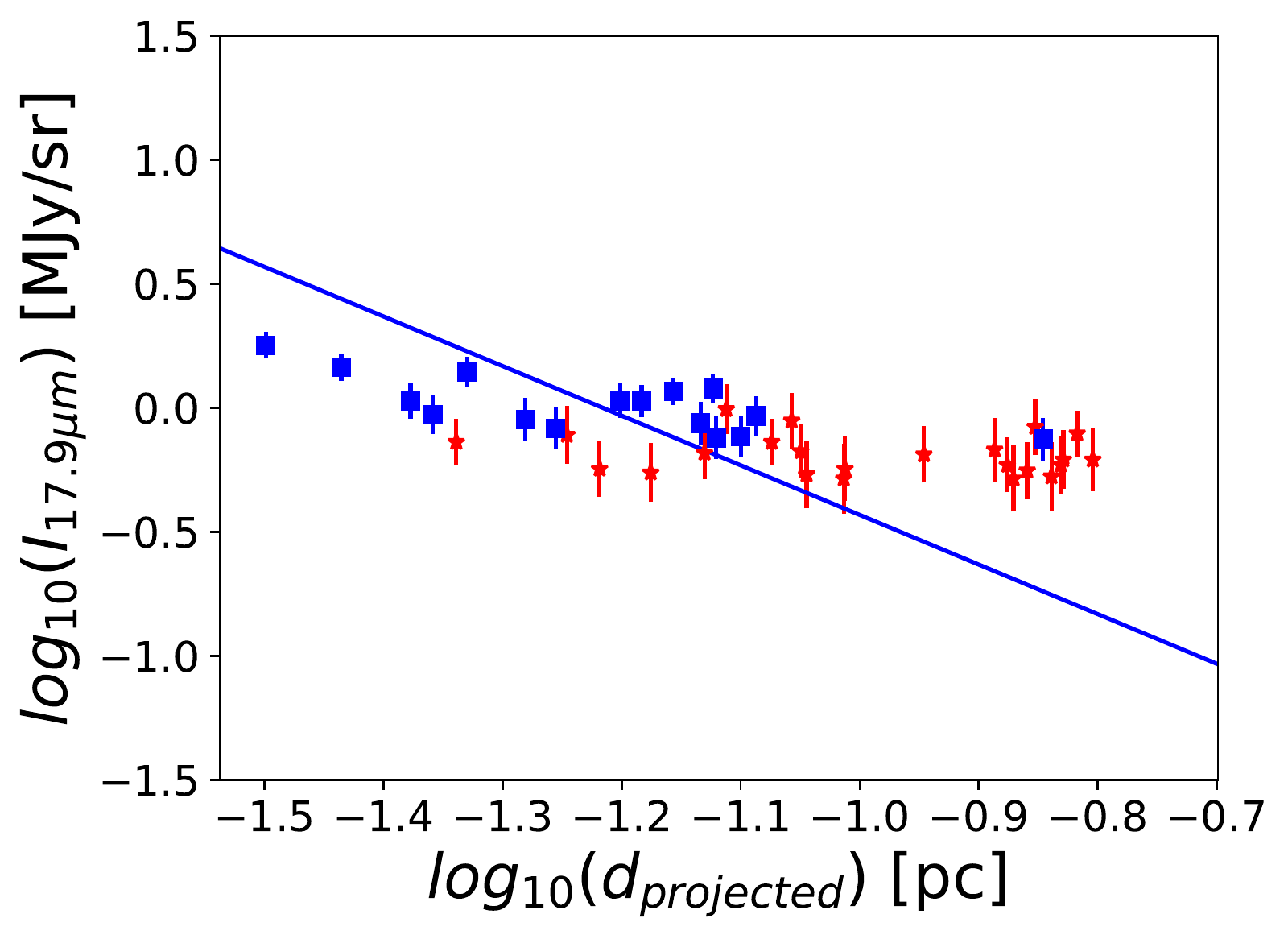} 
	\includegraphics[width=0.32\textwidth]{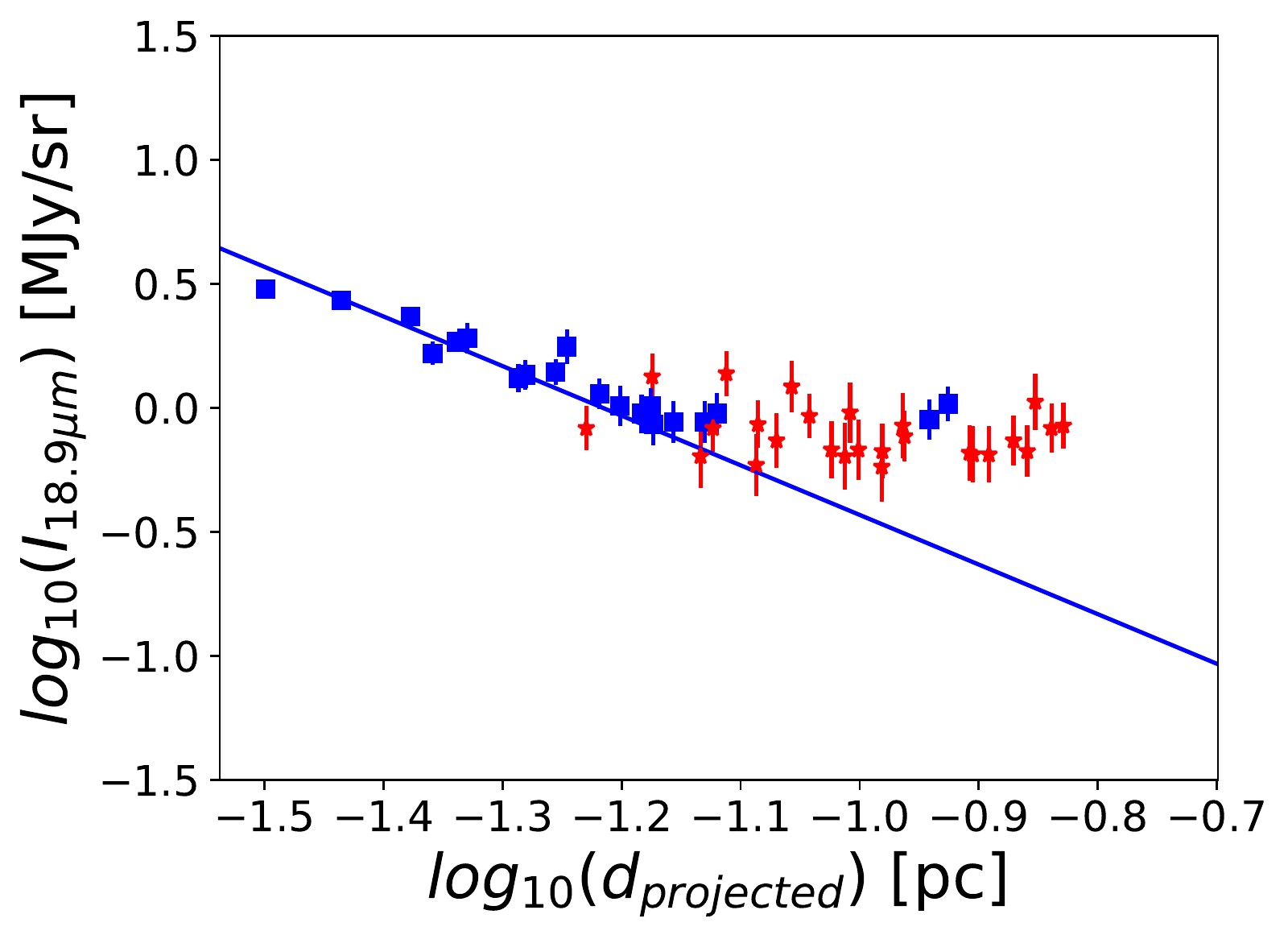} 
	\caption{Logarithmic intensity of selected PAH features with respect to projected distance from the field star, HD\,130079. Blue lines represent the inverse square scaling (with arbitrary normalization), to guide the eye. SN is marked in each by color and symbol with red stars marking the lowest, $3<$\,SN\,$<5$, and blue squares marking all data points with SN\,$>5$. }
\label{Pahtrend1}
\end{figure*}

\section{Results}
\label{sec:results}

\subsection{Trends \& Progressions}
\label{sec:trends}

To a first approximation, one might expect a rather uniform distribution of PAH molecules within the relatively small region covered by the IRS mapping, and therefore that the PAH feature emission reflects, solely, the level of the ionizing continuum radiation. In other words, any spatial variation in the PAH features would be expected to be directly related to the emission of HD\,130079, which is the dominate source of the UV photon flux within the sampled regions of the cloud and as such should decrease with projected distance from the star. However, trends in the feature intensities and intensity ratios observed in DC\,314.8--5.1, appear to be more complex. Below we inspect the individual fitted Drude profiles, along with the integrated PAH features, over distance. In the inspection of features only those with a SNR greater than 3 in at least $20\%$ of the regions are included; here the SN was calculated as, simply, the ratio of the fitted intensity of a given feature over the error in the fit.

First we inspect the individual Drude profile fits from PAHFIT over distance. The resulting trends can be generally divided into two major categories: in particular, as shown in Figure\,\ref{Pahtrend1} and summarized in Table\,\ref{table:pahtrend}, we see either the single ``inverse power-law'' scaling, or the ``inverse power-law + plateau'' trend, i.e. the inverse power-law behavior until a certain distance where the feature flattens out.

\begin{figure}[!th]
\centering
	\includegraphics[width=0.63\columnwidth]{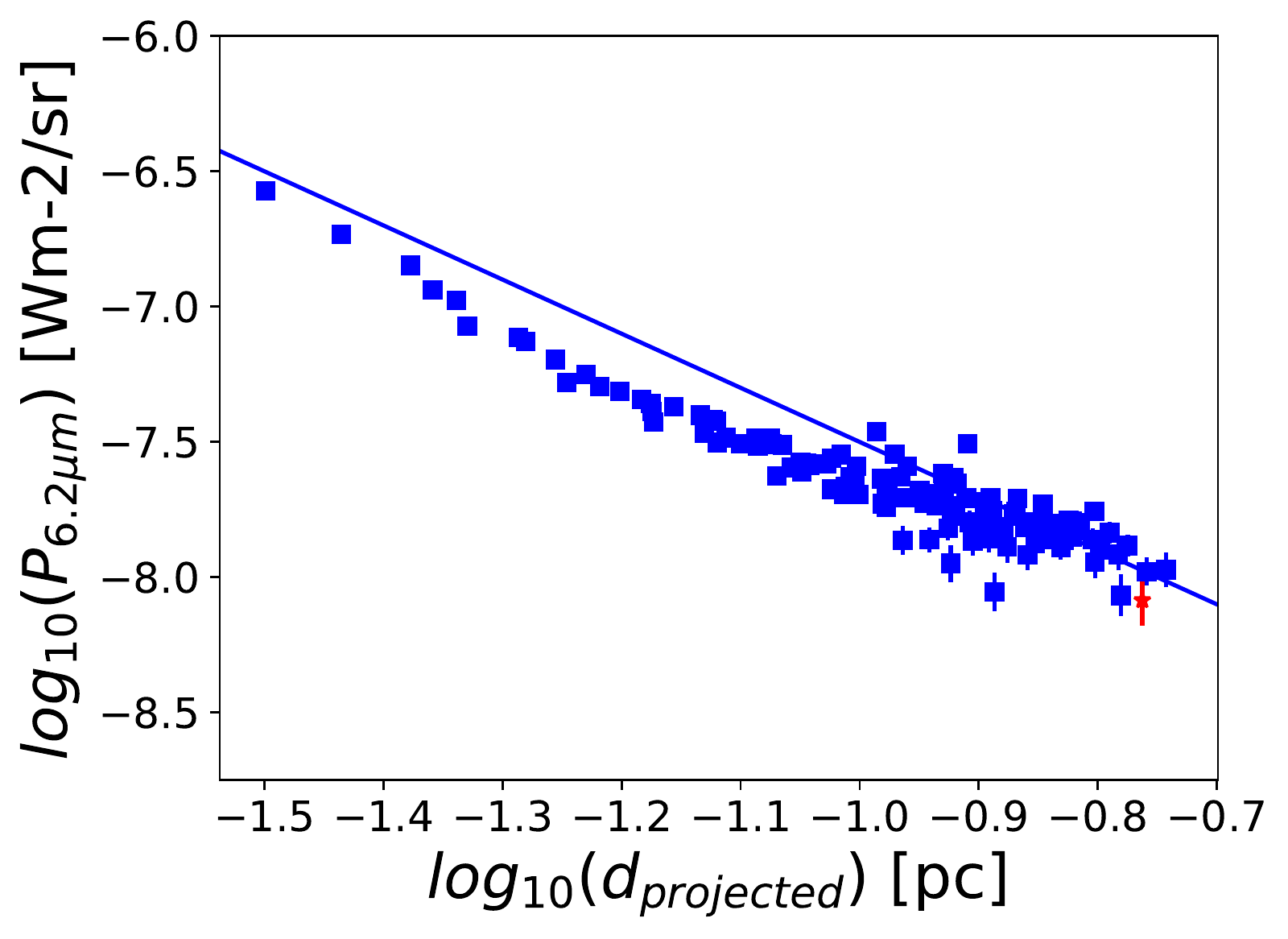} 
	\includegraphics[width=0.63\columnwidth]{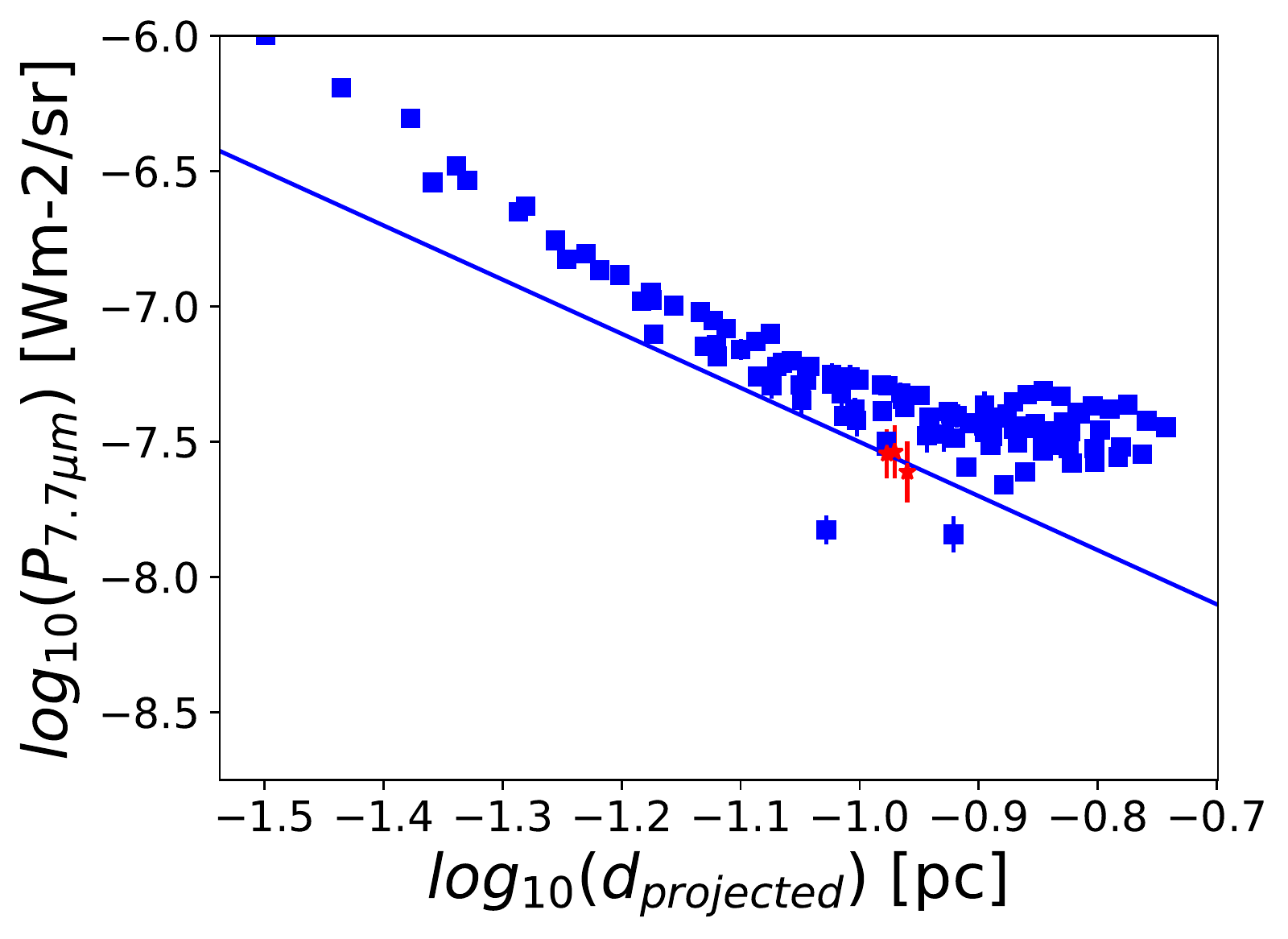} 
    \includegraphics[width=0.63\columnwidth]{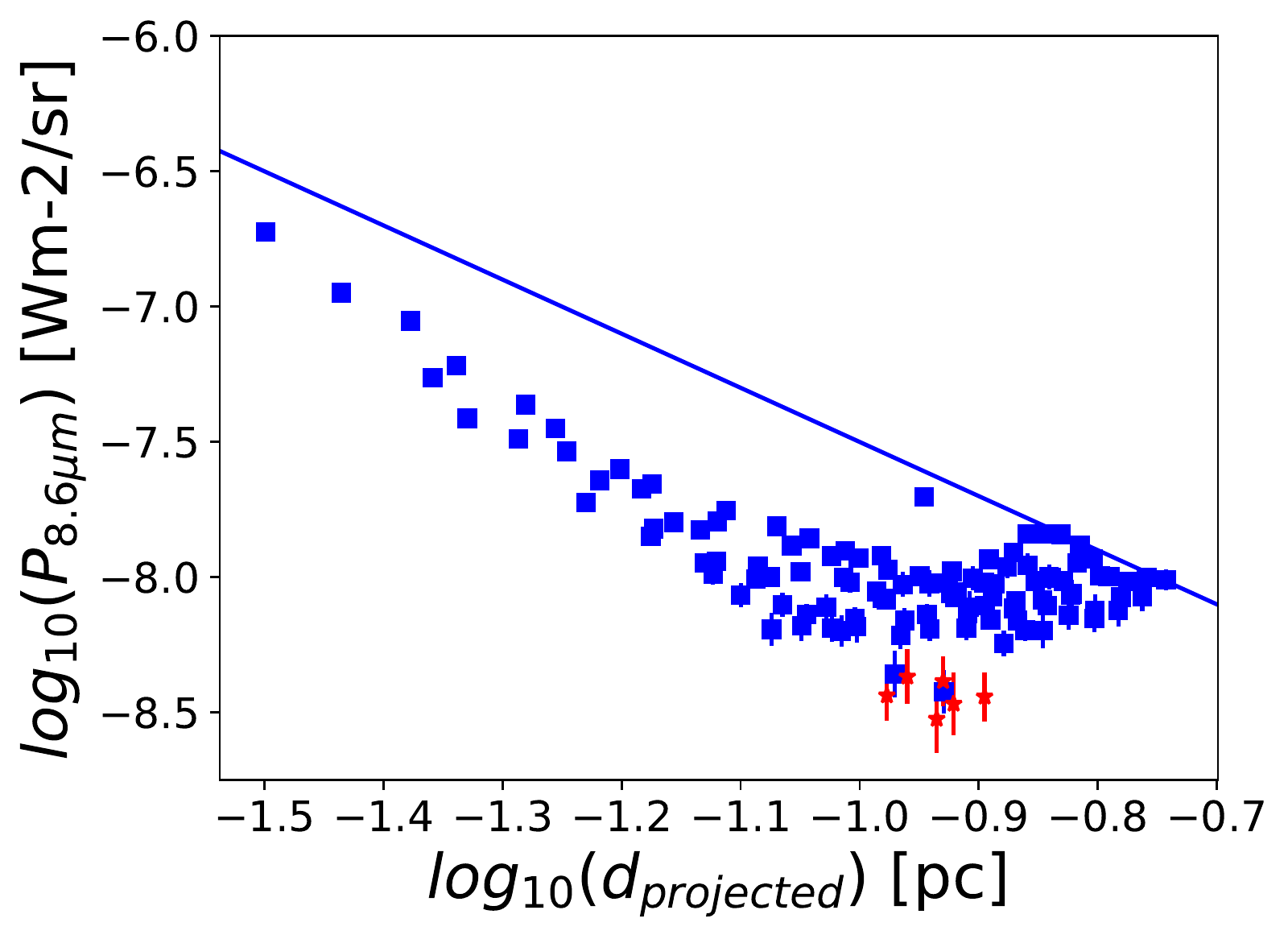} 
	\includegraphics[width=0.63\columnwidth]{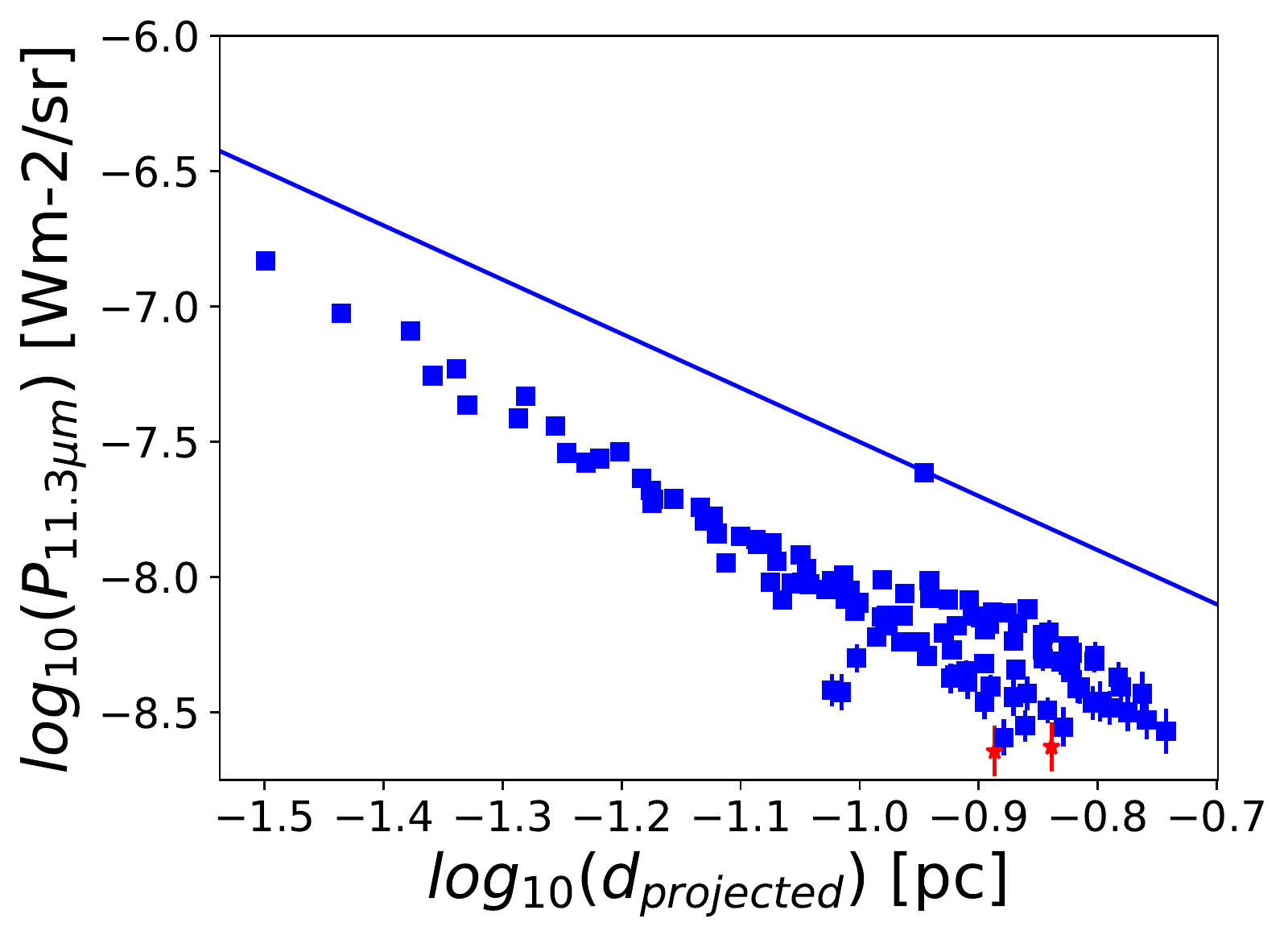} 
	\includegraphics[width=0.63\columnwidth]{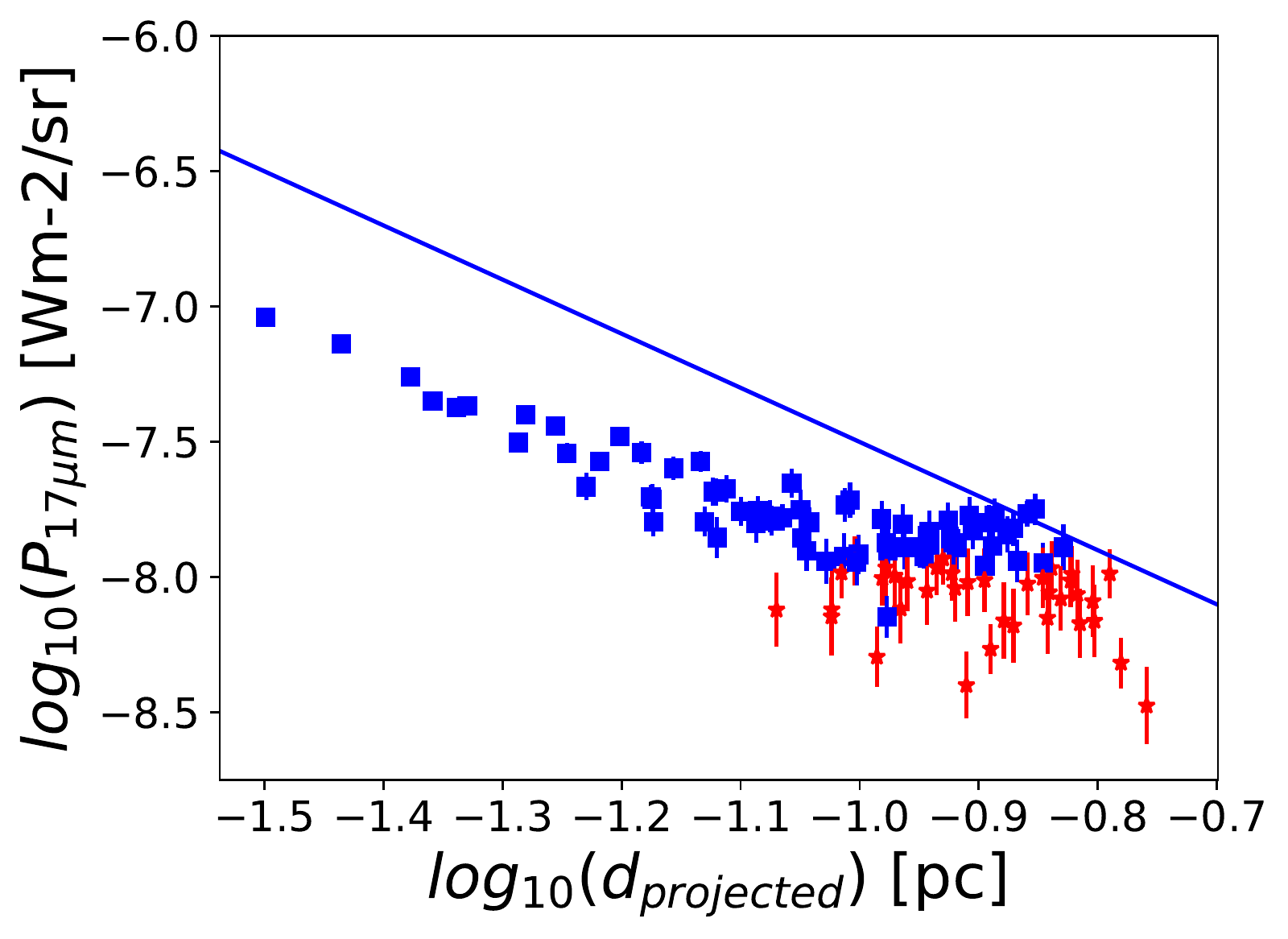}
		\caption{Logarithmic intensity of the integrated power of the main PAH features with respect to projected distance from the field star, HD\,130079. Blue lines represent the inverse square scaling (with arbitrary normalization), to guide the eye. SN is marked in each by color and symbol with red stars marking the lowest, $3<$\,SN\,$<5$, and blue squares marking all data points with SN\,$>5$.}
\label{Mergedtrend}
\end{figure}

The individual PAH features can have multiple Drude profiles contributing to them, and so for such, their integrated strength should be inspected rather than the individual Drude profiles.  This regards in particular five main PAH features, namely 6.2, 7.7, 8.6, 11.3, and 17.0\,$\mu$m, for which the trends are given in the Figure\,\ref{Mergedtrend}. As follows, here the ``power-law'' and ``power-law + plateau''  categories are also represented. Note that for the integrated power trends we consider only the well-defined PAH features, excluding in particular the 8.3\,$\mu$m feature for which PAHFIT returns a high-significant detection in terms of the Drude profile intensity (see Figure\,\ref{Pahtrend1} and Table\,\ref{table:pahtrend}), even though this feature may be rather due to the blending of the neighbouring 7.7 and 8.6 features  \citep[see the discussion in][]{Peeters17}.

\begin{figure*}[!th]
   \centering
    \includegraphics[width=0.87\textwidth]{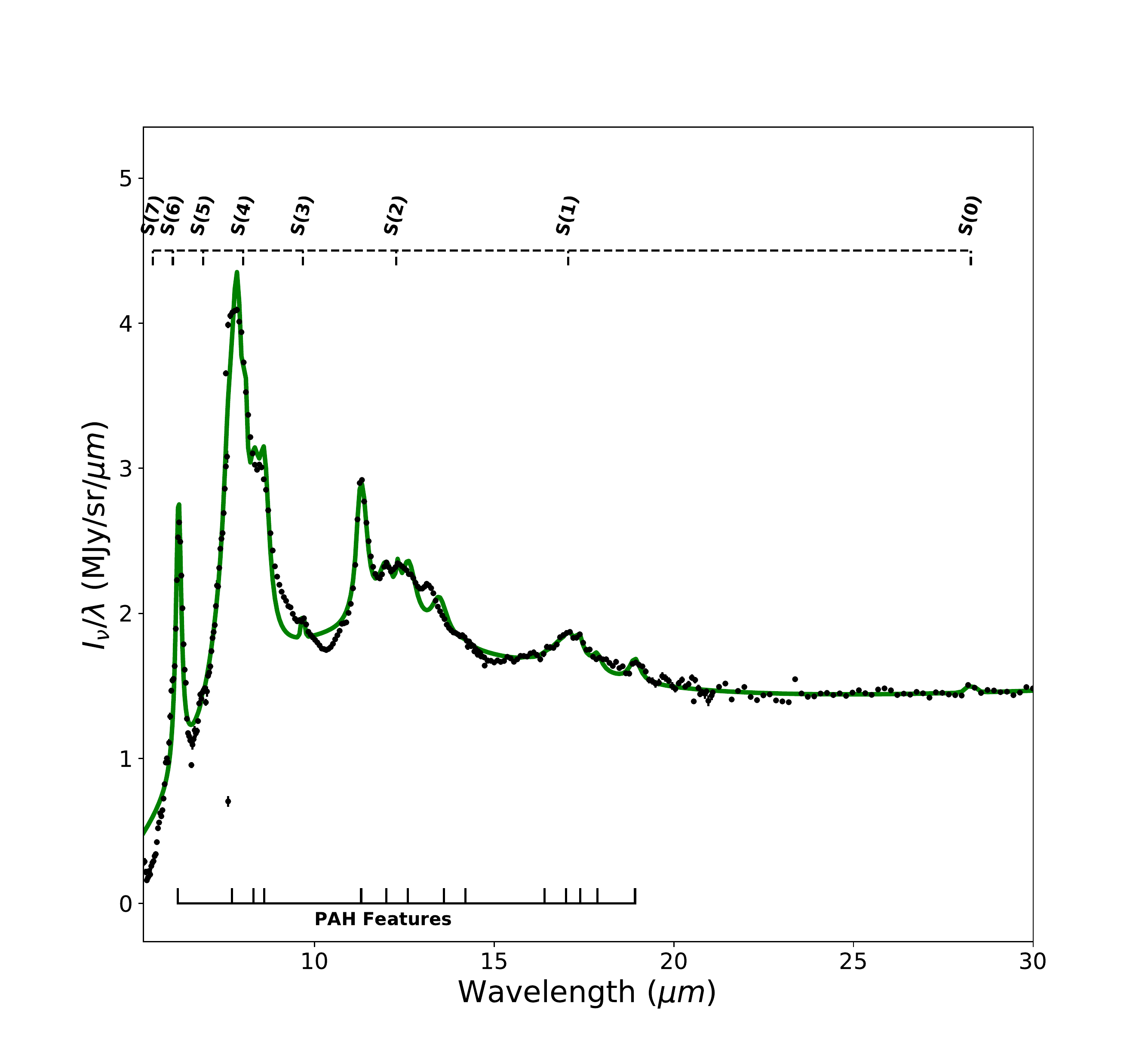}
     \caption{A zoom to the $5.5-30.0$\,$\mu$m segment of the spectrum for the region located at $0.03$\,pc projected distance from HD\,130079 (see the upper panel in Figure\,\ref{Lowres}), highlighting the region encompassing the majority of PAH activity. Green solid line indicates the fully fitted model extracted from PAHFIT \citep{PAHFIT}. Black points mark the real data points with error bars included. Dashed lines at the top of the plot mark the S(0-7) rotational lines of H$_2$ included in the PAHFIT fitting process, solid lines at the bottom of the plot indicate all of the PAH features included in the fitting process (within the $5.5-30.0$\,$\mu$m segment).}
         \label{Zoom1}
\end{figure*}

\subsection{Individual Features and Lines}\label{sec:feat}

\begin{figure*}[!th]
\centering
    \includegraphics[width=0.49\textwidth]{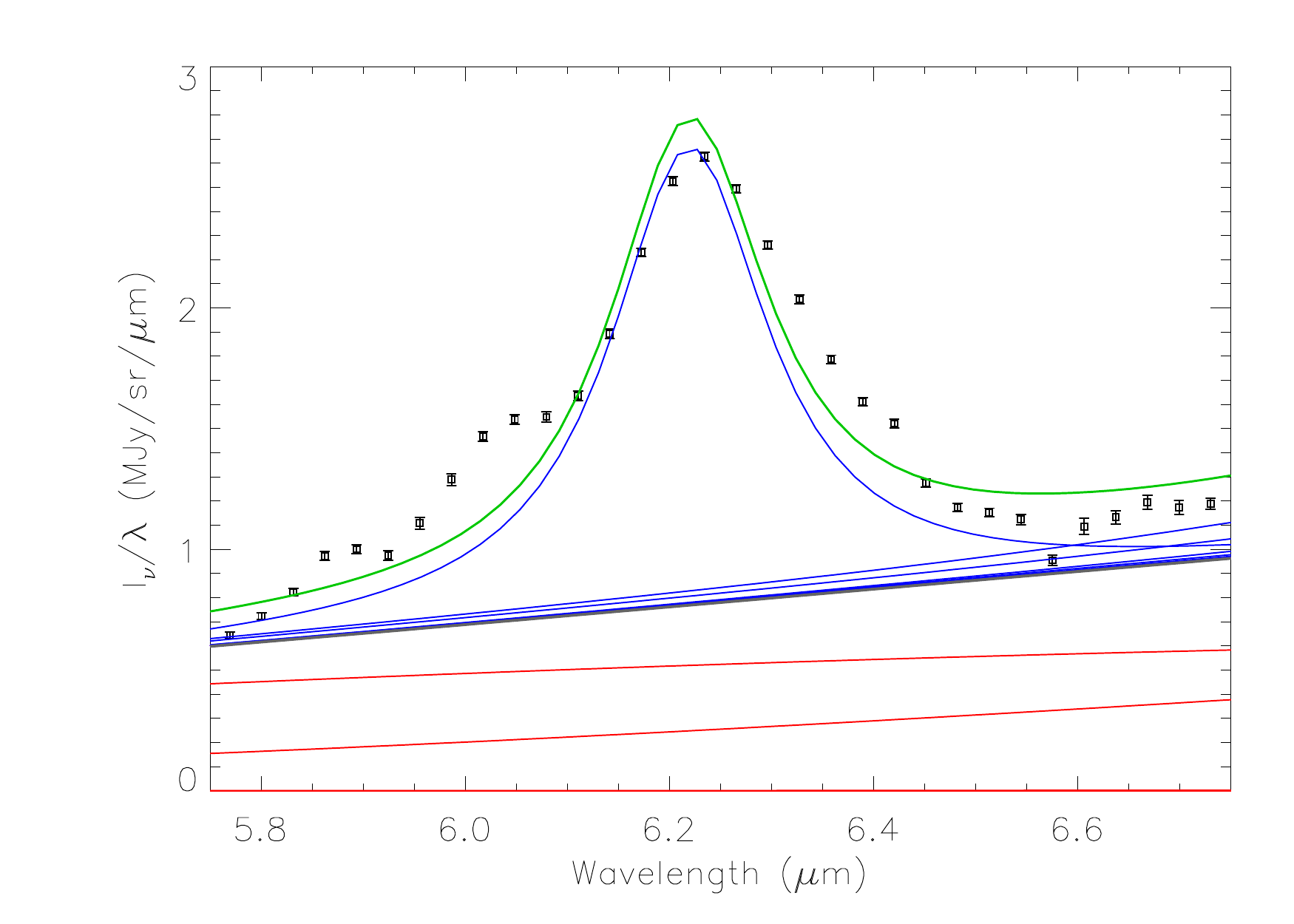} 
    \includegraphics[width=0.49\textwidth]{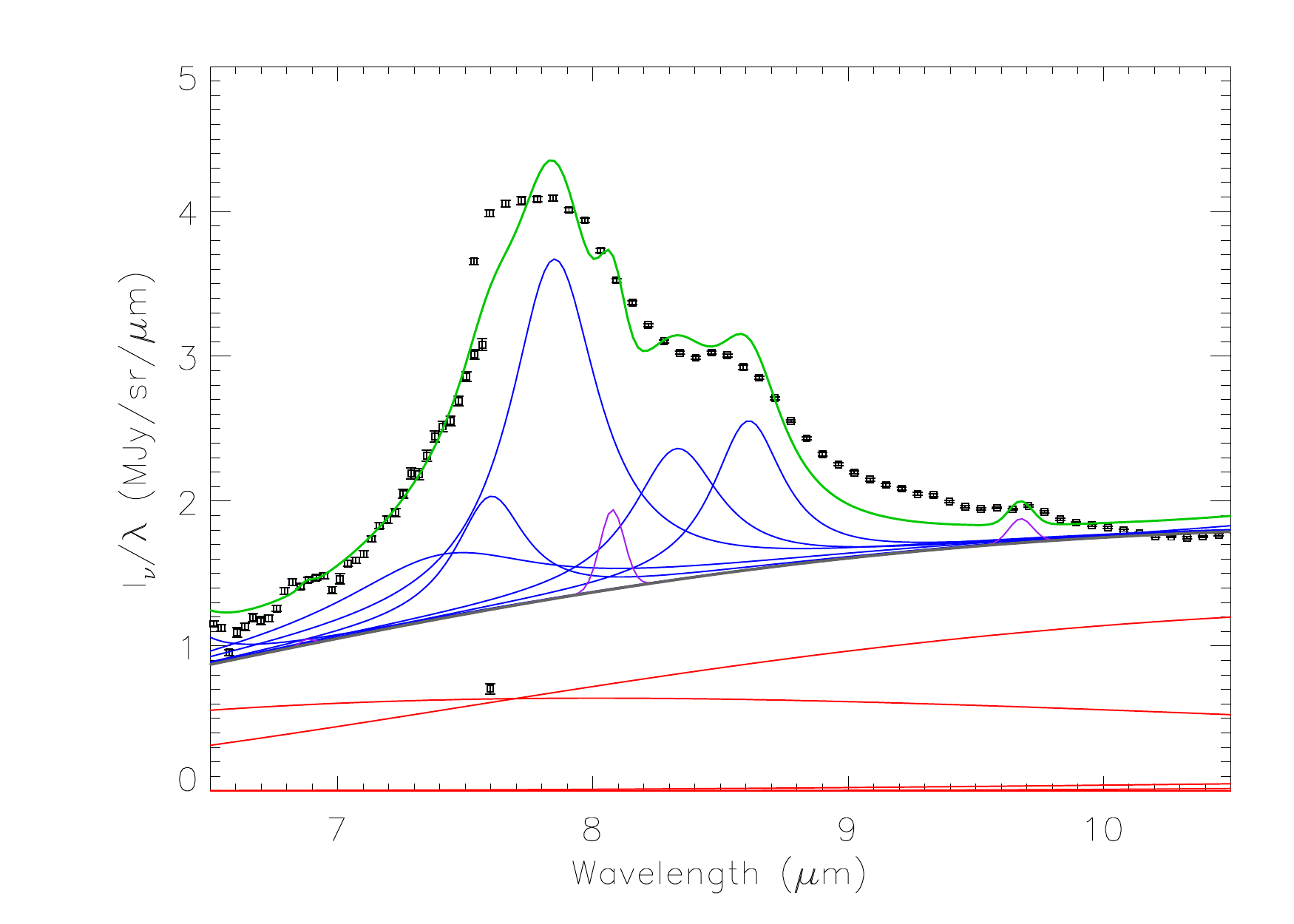} 
    \includegraphics[width=0.49\textwidth]{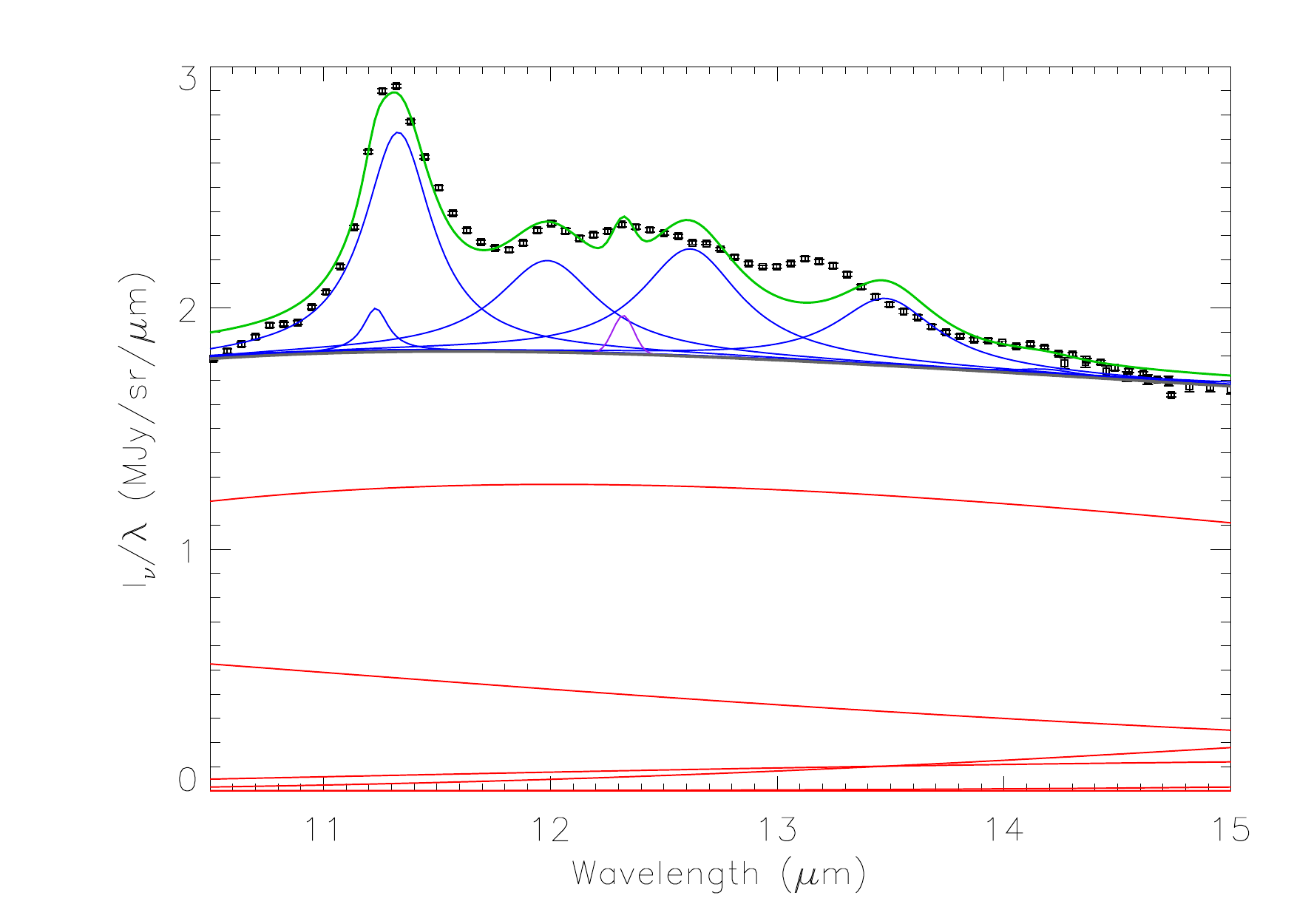} 
    \includegraphics[width=0.49\textwidth]{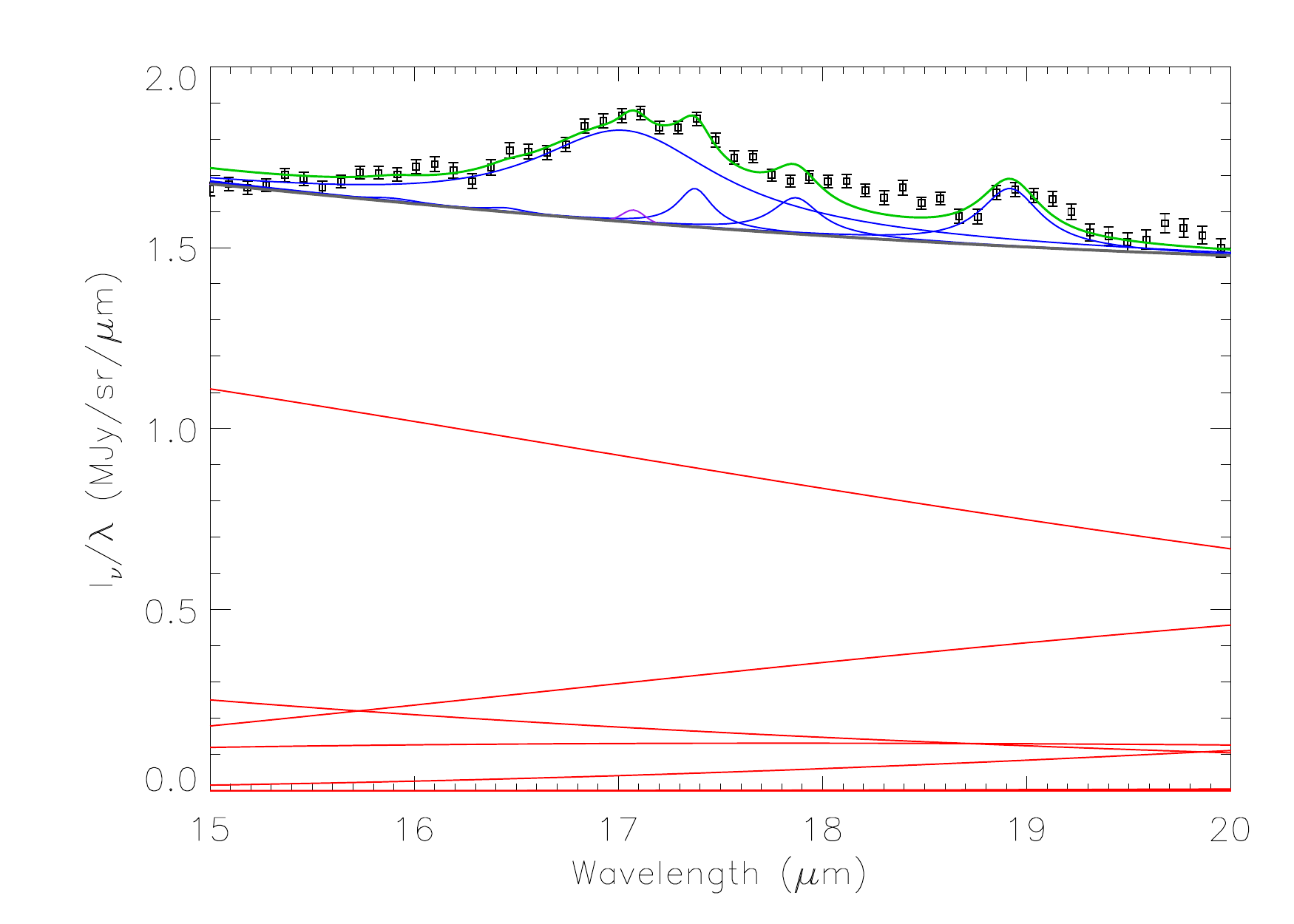} 
      \caption{Zoom in of spectral fit for the region located at $0.03$\,pc projected distance from HD\,130079 (see the upper panel in Figure\,\ref{Lowres}, and also Figure\,\ref{Zoom1}). Lines and points are same as defined in Figure\,\ref{Lowres}, with the green solid line representing the fully fitted model.}
         \label{Zoom2}
   \end{figure*}

Figure\,\ref{Zoom1} presents a zoom to the fully-fitted $5.5-30.0$\,$\mu$m segment of the spectrum for the region located at $0.03$\,pc projected distance from HD\,130079 (see the upper panel in Figure\,\ref{Lowres}), highlighting the region which encompasses the majority of features. As shown, various profiles can contribute to single spectral features, and only by means of a detailed spectral fitting can one disentangle them. In the presentation of individual profiles we adopt the high significance/prevalence criteria, namely, SNR\,$>5$ detections in at least $20\%$ of the regions. Figure\,\ref{Zoom2} presents a more detailed zoom to the more narrow 5.75--6.75, 6.75--10.5, 10--15, and 15--20\,$\mu$m segments of the spectrum for the same region, including model curves denoting various lines identified during the fitting procedure. The same decomposition was applied to all the analyzed 117 regions, and the results are given below in the following subsections. 

\subsubsection{5--7 Micron Features}
\label{sec:5-7}

In the upper left panel of Figure\,\ref{Zoom2}, the decomposition of the  $6.2$--$6.3$\,$\mu$m feature is shown. Our system has a consistent peak wavelength across the 117 regions of $6.22$\,$\mu$m. \citet{Peeters02} perform an investigation of the shift of the $6.3$\,$\mu$m PAH feature due to the addition of a nitrogen atom in the molecule. The $6.3$\,$\mu$m feature is represented by pure PAHs where as a shifted peak around $6.2$\,$\mu$m is emitted by substituted PAH species. Our observations are consistent with the findings presented in \citet{Peeters02} for reflection nebulae, in which, it is suggested that the shifting of the peak wavelength is due to the presence of nitrogen in the observed PAH molecules. 

Another factor that can influence the positioning of the 6.2 peak, is the presence of ice absorption around 6\,$\mu$m from water ice. The ice features at 6\,$\mu$m can be confused when significant PAH features are present at 5.25, 5.7, 6.2, and 7.7\,$\mu$m \citep[][and refs. therein]{Peeters02}. Indeed we see strong features at 6.2 and 7.7\,$\mu$m due to our PAH compounds, which makes it difficult to diagnose any potential ice absorption around 6\,$\mu$m. 

The lack of water ice in this spectral range is consistent with what is expected for the cloud. The observations indicate that the ice formation threshold occurs for $A_V$ above 1.6\,mag toward different regions \citep[e.g.,][]{Whittet01,Chiar11,Boogert13,Whittet13}. Nevertheless, the visual extinction at the edges of DC\,314.8-5.1 is estimated to be equal to $1.08\pm 0.2$\,mag, without the foreground effect \citep{Whittet07}. Other molecules that show strong IR absorption features at around 6\,$\mu$m, such as HCOOH, H$_2$CO, and CH$_3$CHO, are more volatile than water, and thus would be formed in regions with larger extinction values ($A_V > 3$\,mag). 

\subsubsection{7--10 Micron Features}

The 7--10\,$\mu$m spectral region is complex and is decomposed into several Drude profiles.  The statistically significant and well defined PAH features in this spectral selection are the ones centered at 7.7 and 8.6 \,$\mu$m, as shown in Figure\,\ref{Zoom2} upper right. The $7.7$\,$\mu$m profile is fitted with three Drude profiles centered at $7.4$, $7.6$, and $7.8$\,$\mu$m. The peak of this feature is located longwards of $7.6$\,$\mu$m, which suggests a more dominant $7.8$\,$\mu$m component. According to \citet{Peeters02}, this behaviour is common among reflection nebulae, which typically display, in addition, the shifted $6.2$\,$\mu$m PAH feature, as is the case in DC\,314.8--5.1 (see Section\,\ref{sec:5-7}).

 As discussed in \citet{Bregman05}, PAHs in the diffuse ISM that have had long exposure to UV emission, are dominated by a $7.7$\,$\mu$m emission feature with a peak around $7.6$\,$\mu$m. However, after integration into a cloud, these PAHs are chemically processed, and so the emission feature becomes shifted towards $7.8$\,$\mu$m, and then back toward $7.6$\,$\mu$m only when having been exposed to an ionizing UV continuum in a reflection nebulae. The fact that, in the analyzed segment of DC\,314.8--5.1, we see a dominant emission component centered near $7.8$\,$\mu$m, with a non-negligible contribution from the emission at $7.6$\,$\mu$m, implies therefore, that in addition to the ``cloud-native'', chemically processed, PAH population, we also see a fraction of PAHs already being processed by the UV radiation of the neighbouring star, HD\,130079.

\subsubsection{10--15 Micron Features}

A strong $11.2-11.3$\,$\mu$m feature is commonly attributed to neutral PAHs within the cloud, and is correlated with many other PAH features. The profile in PAHFIT is comprised of two Drude profiles located at $11.2$ and $11.3$\,$\mu$m. Figure\,\ref{Zoom2} bottom left shows that the $11.2$ and $11.3$ Drude profiles contribute to one PAH feature with a peak wavelength of $\sim$\,$11.25$, but with a significantly more dominant contribution from the $11.3$\,$\mu$m profile, and a lesser contribution from the $11.2$\,$\mu$m profile.  As is visible in Figure\,\ref{pahdb}, this feature is dominated by large primarily neutral PAHs with the red wing of the feature being contributed to by anion PAHs. The $11.3$\,$\mu$m Drude profile follows the inverse power-law trend nicely, unlike the $11.2$\,$\mu$m Drude profile. The overall integrated feature, however, does follow the power-law trend.

We note that PAHFIT does not fully converge in the $12-14$\,$\mu$m region. Although the $12$\,$\mu$m and $12.6$ PAH features are statistically significant, it is visible in Figure\,\ref{Zoom2}, that these lines are included in an attempt to fit the plateau present in this region and are not clear emission line detections.  In \citet{Peeters17}, this plateau ($10-15$\,$\mu$m) is attributed to a blending of irregular, small, and clustered PAHs emitting from the C-H bending modes \citep[single, duo, trio, and quartet H atoms; see][for an in-depth discussion of these plateau features]{Tielens08,Peeters17,Allamandola89,Bregman89}. Due to PAHFIT not taking into account the plateau nature of this spectral section, we do not further consider the features between 12--14\,$\mu$m, as we cannot be confident with these detections.

\subsubsection{15--20 Micron Features}

In the region past 15\,$\mu$m, see Figure\,\ref{Zoom2} lower right panel, PAHFIT takes into account six Drude profiles at 15.9, 16.4, 17.0, 17.4, 17.9, and 18.9\,$\mu$m to fit the PAH features. The prominent (and as such statistically significant) features present here are the integrated PAH feature at 17.0\,$\mu$m as well as the PAH Drude profile at 18.9\,$\mu$m.

\subsection{Dust Continuum}
\label{sec:dust}

\begin{figure}[!th]
 \centering
     \includegraphics[width=\columnwidth]{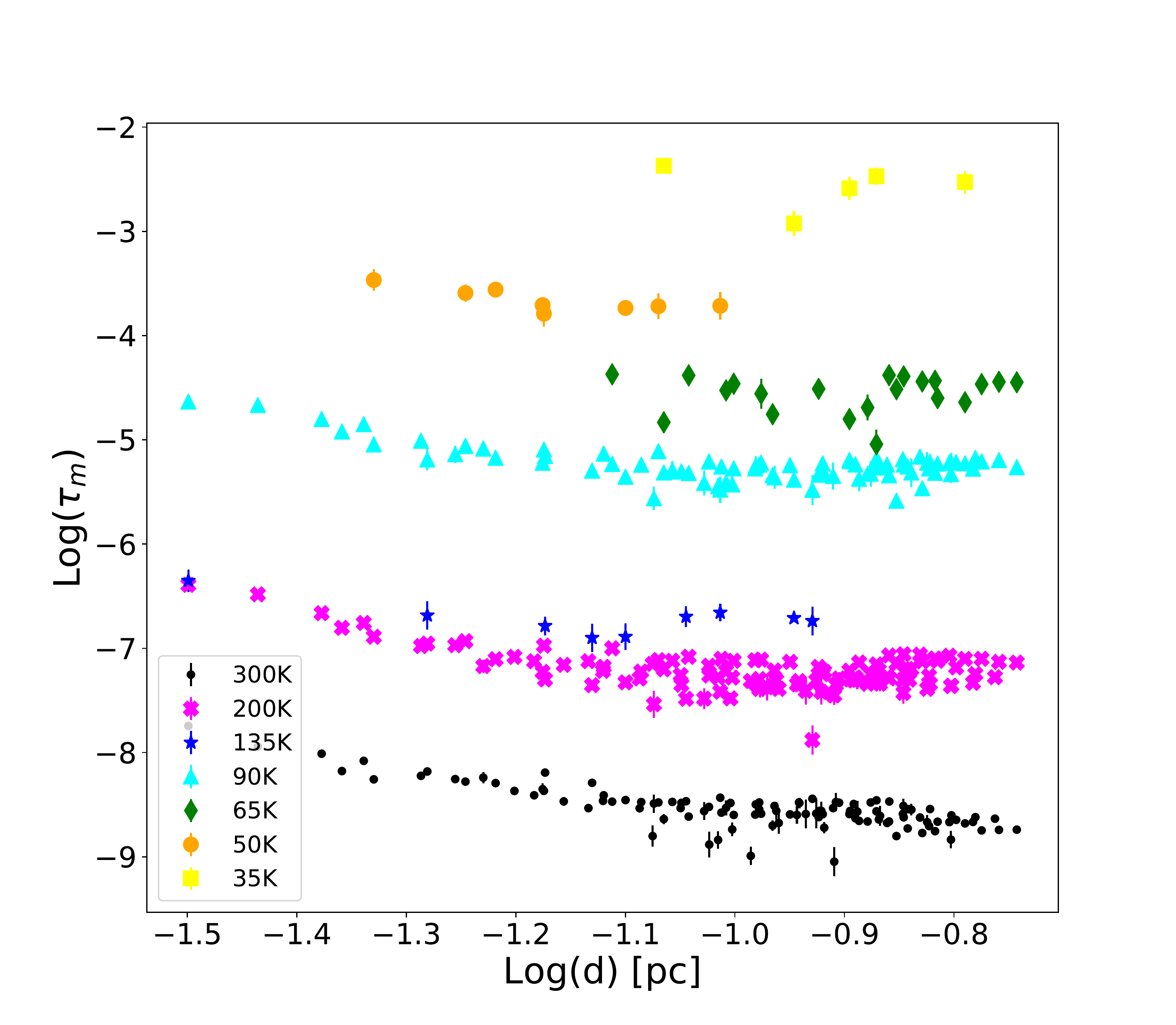} 
   \caption{The relative strength of the dust continuum components with fixed model temperatures of 300, 200, 135, 90, 65, 50, 40, and 35\,K, with increasing (projected) distance from the star HD\,130079. }
         \label{Temp}
\end{figure}

The PAHFIT process allows for eight thermal dust continuum components (featureless modified blackbodies), at fixed temperatures in the range 35--300\,K; the set of best-fit amplitudes of all the components is expected to closely represent a smooth distribution of grain temperatures within the source \citep{PAHFIT}. Based on the fits obtained for each region extracted from the Spitzer IRS mapping data for DC\,314.8--5.1, we can investigate the dominance of the eight model dust continuum components fit, with respect to the distance from the star, see Figure\,\ref{Temp}. Data points with a SNR\,$<3$  were excluded, which comprised many of the lower temperature components. SN was taken from the fitted profile and the respective error in that fit.

Firstly, we note that not all the dust continuum components are required by the fitting procedure in all the 117 regions analyzed. The prevalence of the dust components in the regions varies widely from $\sim5\%$ (e.g., the 35\,K and 40\,K components), up to even $>95$\% in the case of the 200, and 300\,K components. This indicates, that the continuum emission of the dust, with the provided model temperatures, requires a full mix of lower and higher temperature blackbodies in the majority of sampled regions for a comprehensive fit of the continuum emission, in agreement with the findings by \citet{Whittet07}.

In Figure\,\ref{Temp}, we plot the extracted relative normalization values for each temperature bin, $\tau_m$\,, as a function of the distance from the star. These relative normalizations are defined in PAHFIT through the equation for the dust continuum intensity
\begin{equation}
    I_{\nu}^{\rm (dust)} = \sum_{m=1}^{8}\tau_m \, \frac{B_\nu (T_m)}{(\lambda/\lambda_0)^2}
\end{equation}
where $B_\nu$ is the blackbody function, $T_m$ are the selecetd thermal dust continuum temperatures, and $\lambda_0 = 9.7$\,$\mu$m \citep{PAHFIT}. In the figure, each temperature bin is color coordinated with their respective points for the fit shown. 

\begin{figure*}[ht]
   \centering
   \includegraphics[width=0.49\textwidth]{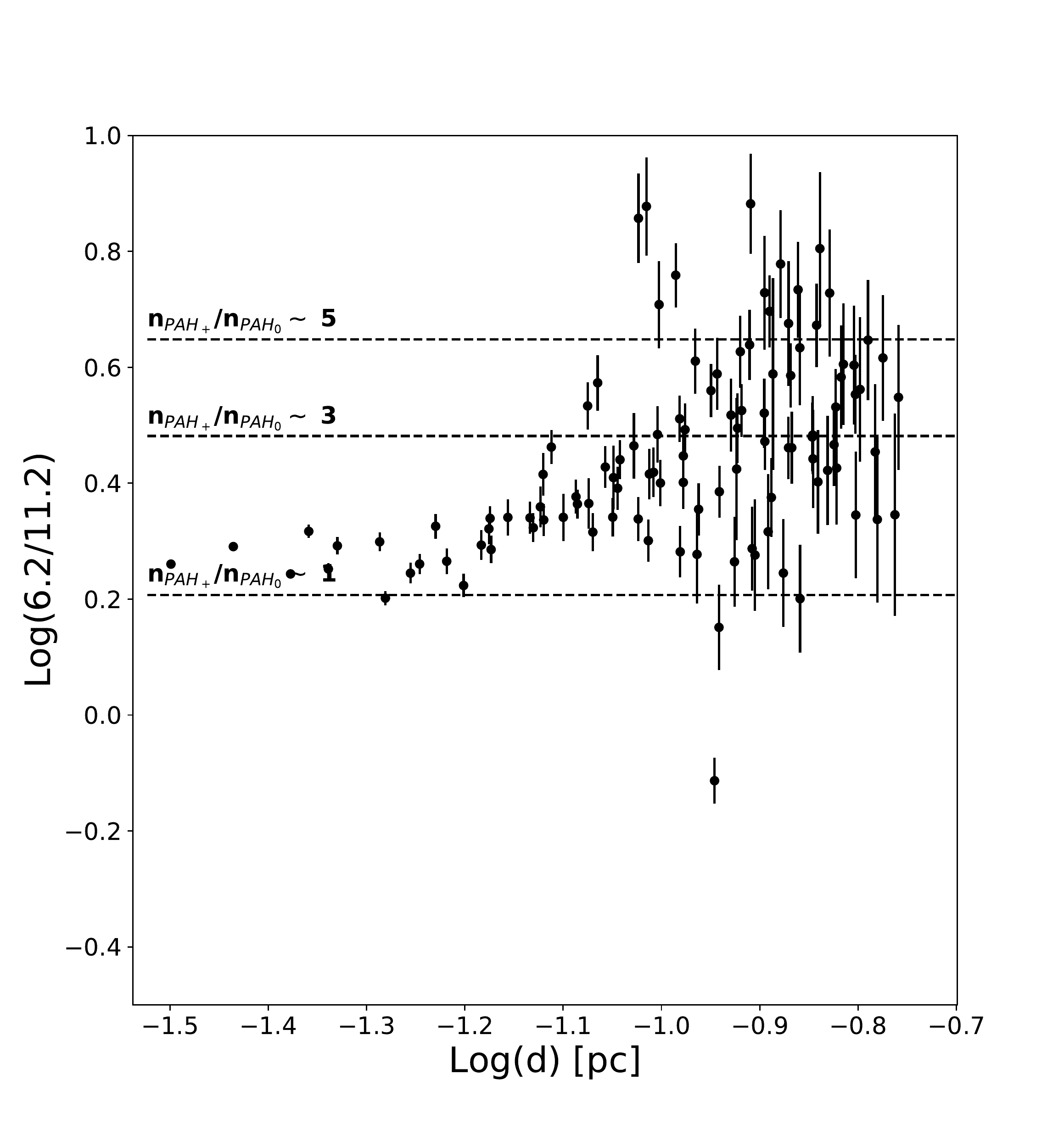}
   \includegraphics[width=0.49\textwidth]{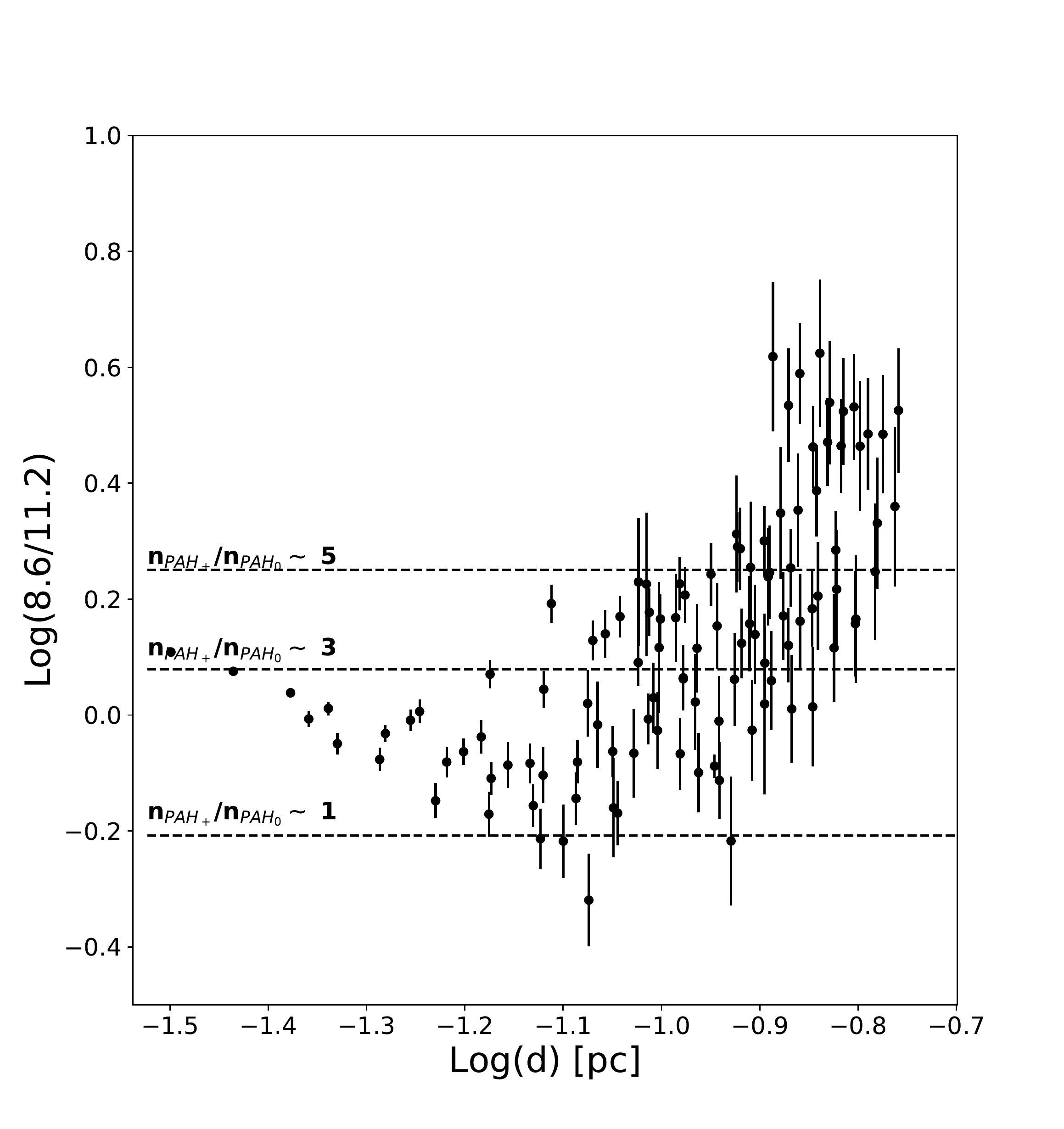}
   \includegraphics[width=0.49\textwidth]{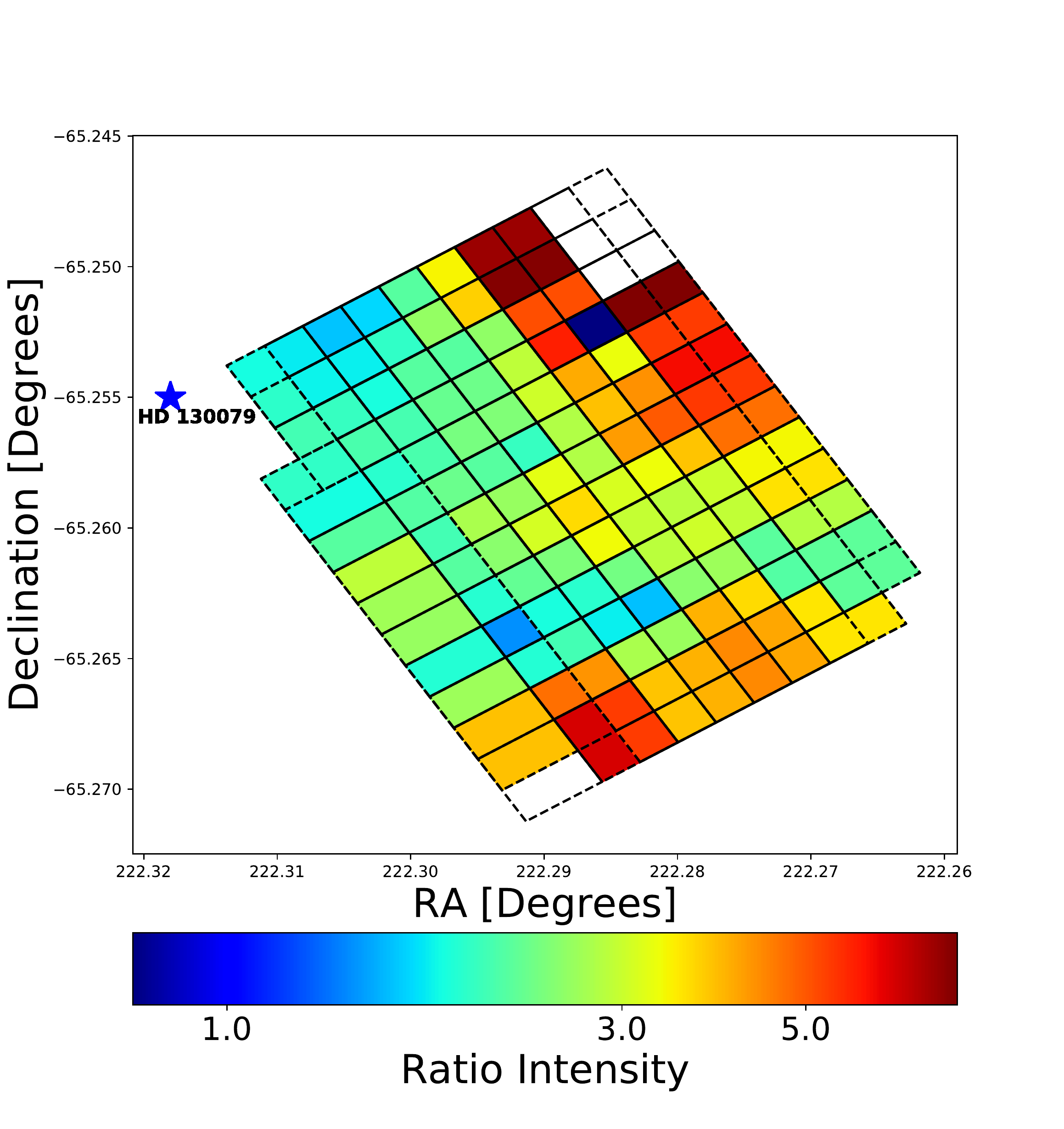}
   \includegraphics[width=0.49\textwidth]{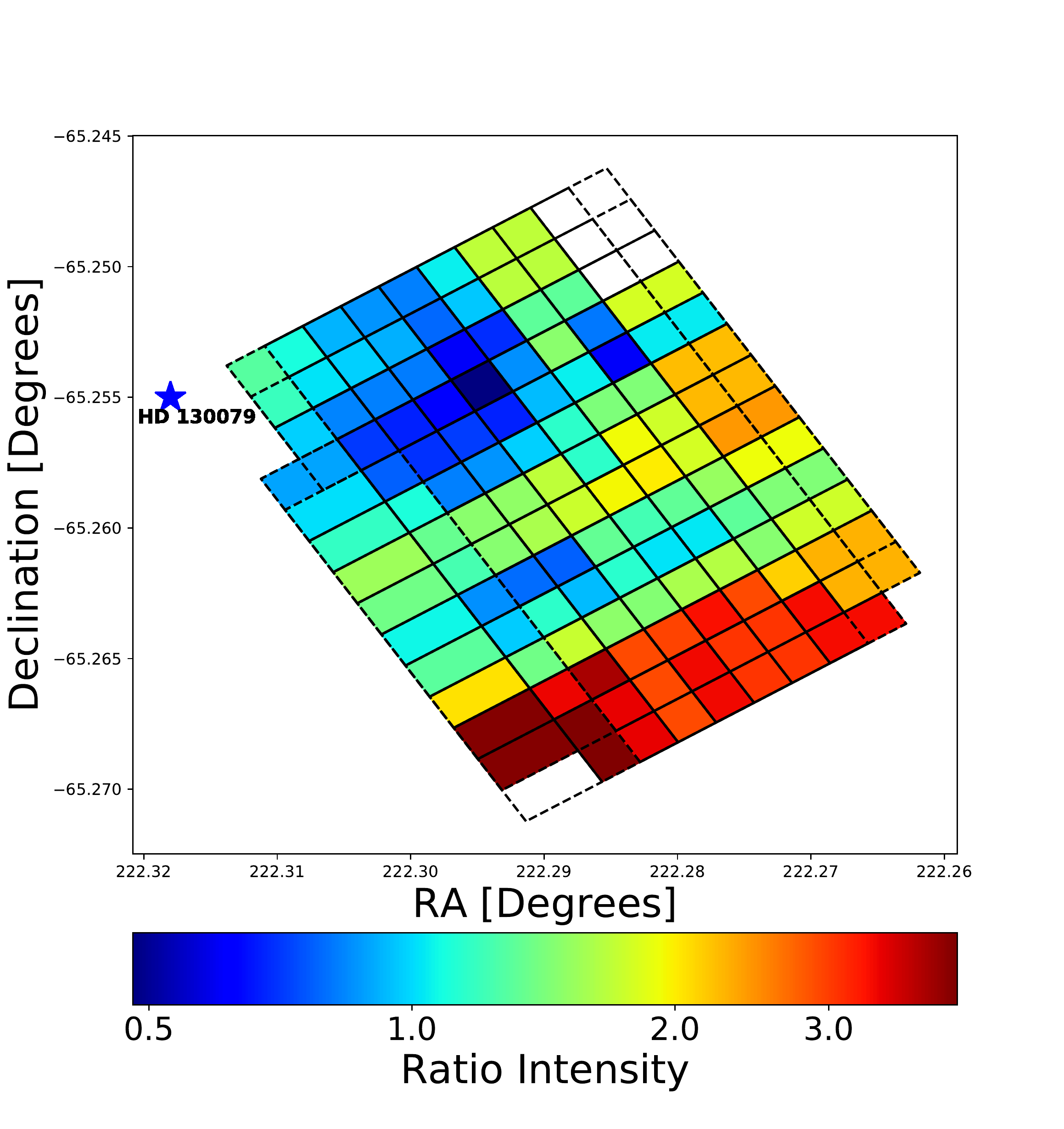}
   \caption{(top panels) Diagram showing the 6.2/11.2 intensity ratio (left) and 8.6/11.2 intensity ratio (right), based on the integrated profiles, versus the projected distance from HD\,130079. Horizontal lines correspond to the intensity ratios expected for given values of the ionization parameter $n_{\rm PAH+}/n_{\rm PAH0}$, based on the best-fit correlations by \citet{Boersma18}, namely $I_{6.2 \mu m}/I_{11.2 \mu m} = 0.90 +0.71 \times (n_{\rm PAH+}/n_{\rm PAH0})$, and $I_{8.6 \mu m}/I_{11.2 \mu m} = 0.33 +0.29 \times (n_{\rm PAH+}/n_{\rm PAH0})$. (bottom panels) RA/DEC maps displaying the 6.2/11.2 intensity ratio (left) and 8.6/11.2 intensity ratio (right).}
         \label{Ionization}
   \end{figure*}

Studies have shown that quiescent, non-active, cores of dust clouds, have on average temperatures around 7--20\,K, with colder temperatures dominating in the central regions \citep{Bergin07}. Molecular clouds, on the other hand, maintain on average temperatures around 20--30\,K \citep{Ward02}. In the case of DC\,314.8--5.1, the most prominent dust continuum component corresponds to the temperature of 40 and 35\,K, having orders of magnitude higher normalizations than the hotter counterparts. These are, however, the lowest temperature bins considered in PAHFIT, and no lower-temperature dust components are considered during the fitting because the dust below 35\,K is too cold to make any contribution at $\lambda < 40$\,$\mu$m \citep{PAHFIT}.

\section{Discussions}
\label{sec:discussion}

 \subsection{PAH Ionized Fraction}
 \label{sec:ionized}
 
\citet{Boersma14} derived a relation between the number density ratio of PAH cations to neutrals, $n_{\rm PAH+}/n_{\rm PAH0}$, and the main physical parameters in the system, including gas temperature $T$, ionizing radiation field $G_0$, free electron number density $n_e$, and the PAH number of carbon atoms $N_{\rm C}$, namely
\begin{equation}
    \frac{n_{\rm PAH+}}{n_{\rm PAH0}} \simeq 5.11 \times 10^{-6} \, N_{\rm C}^{1/2} \, G_0 \, T^{1/2} \, n_e^{-1} \, ,
    \label{eq:ratio}
\end{equation}
where $T$ is given in Kelvin, $G_0$ is in the Habing field units, and $n_e$ is in cm$^{-3}$. Note that the PAH ionization parameter is often defined as $\chi \equiv G_0 \, T^{1/2} \, n_e^{-1}$, so that for the exemplary $N_{\rm C} \simeq 50$ one has $(n_{\rm PAH+}/n_{\rm PAH0}) \sim 0.37 \, (\chi / 10^{4})$.
 
\begin{figure*}[!th]
\centering
	\includegraphics[width=0.49\textwidth]{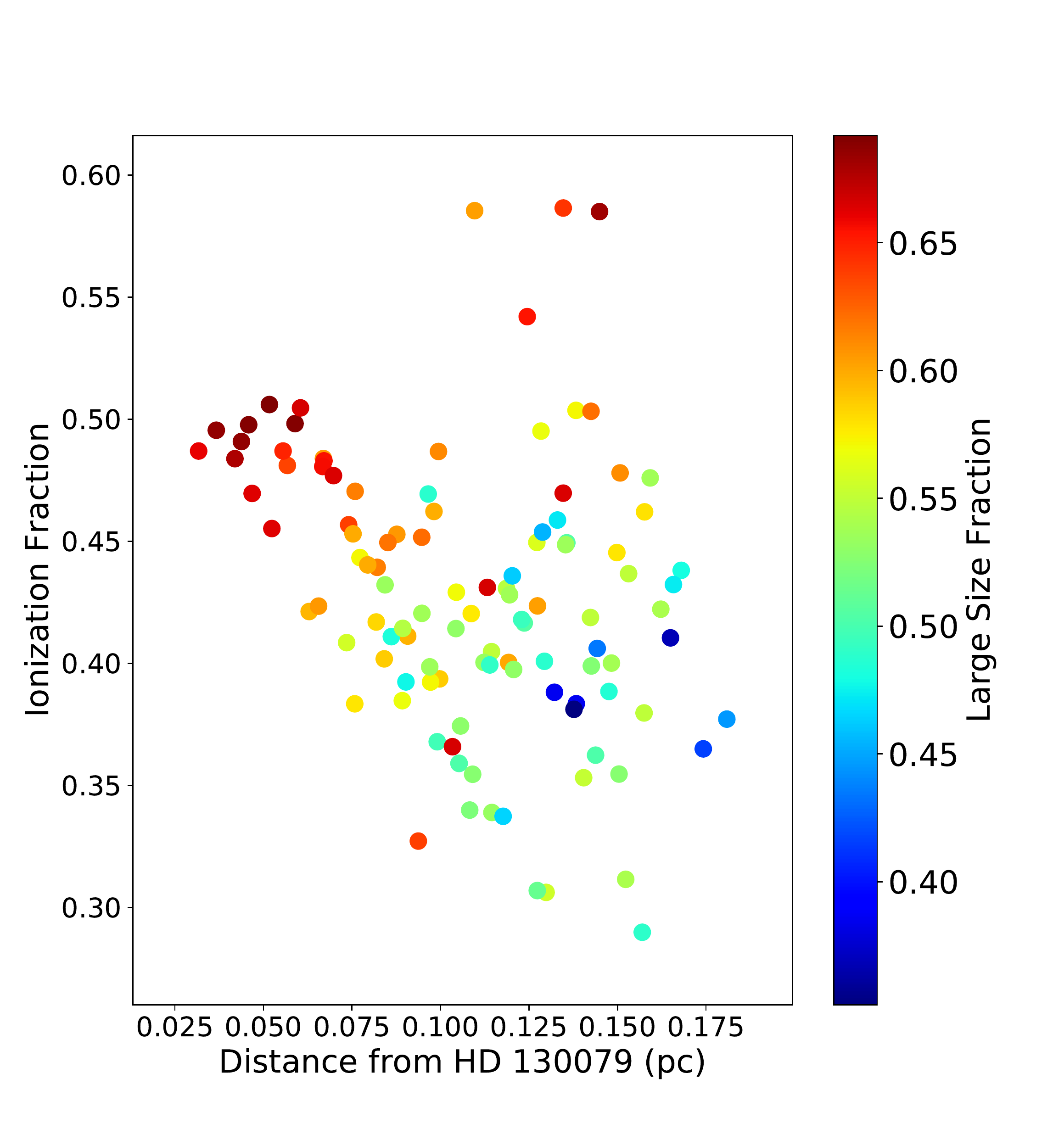}
	\includegraphics[width=0.49\textwidth]{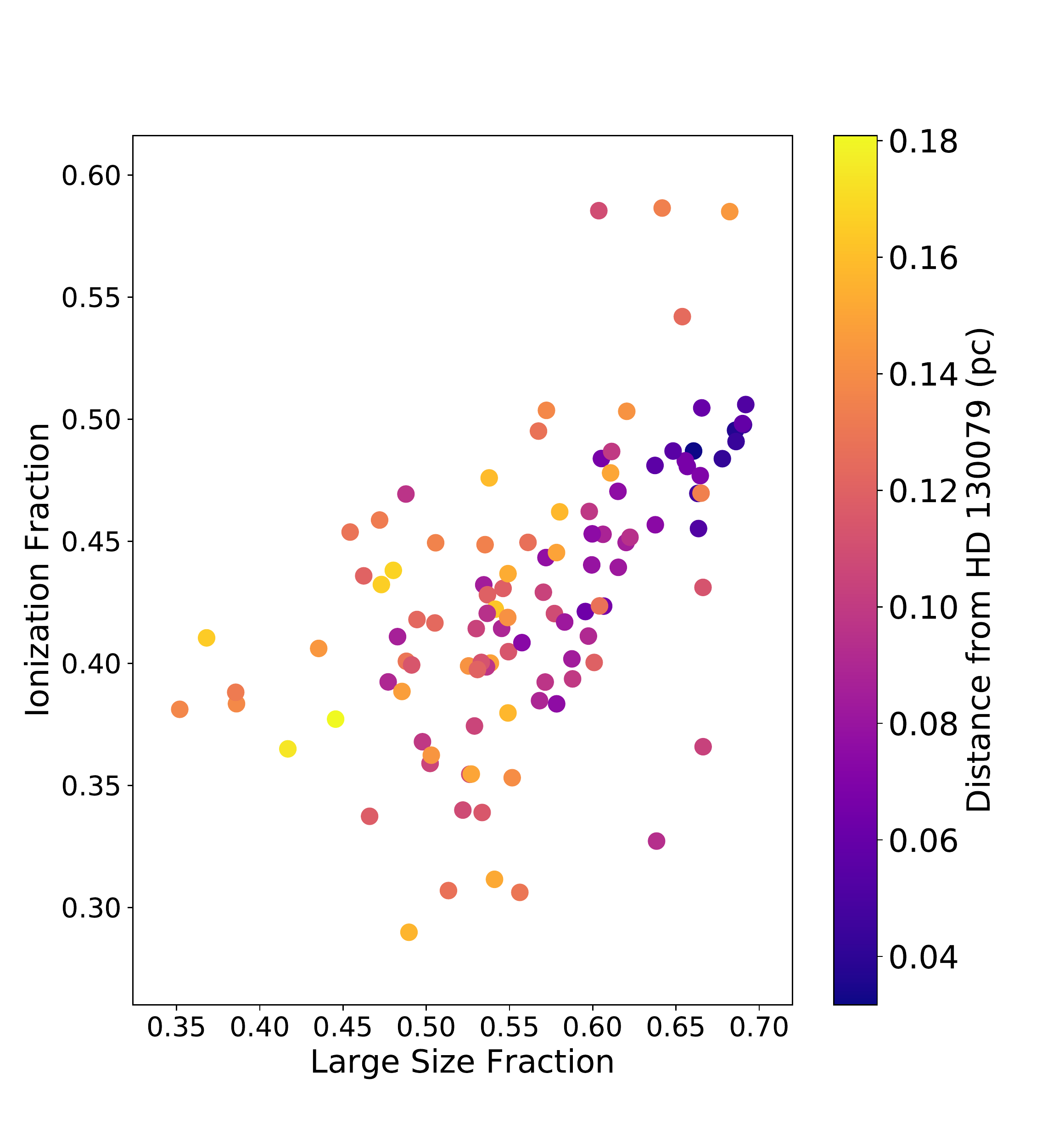}
	\includegraphics[width=0.49\textwidth]{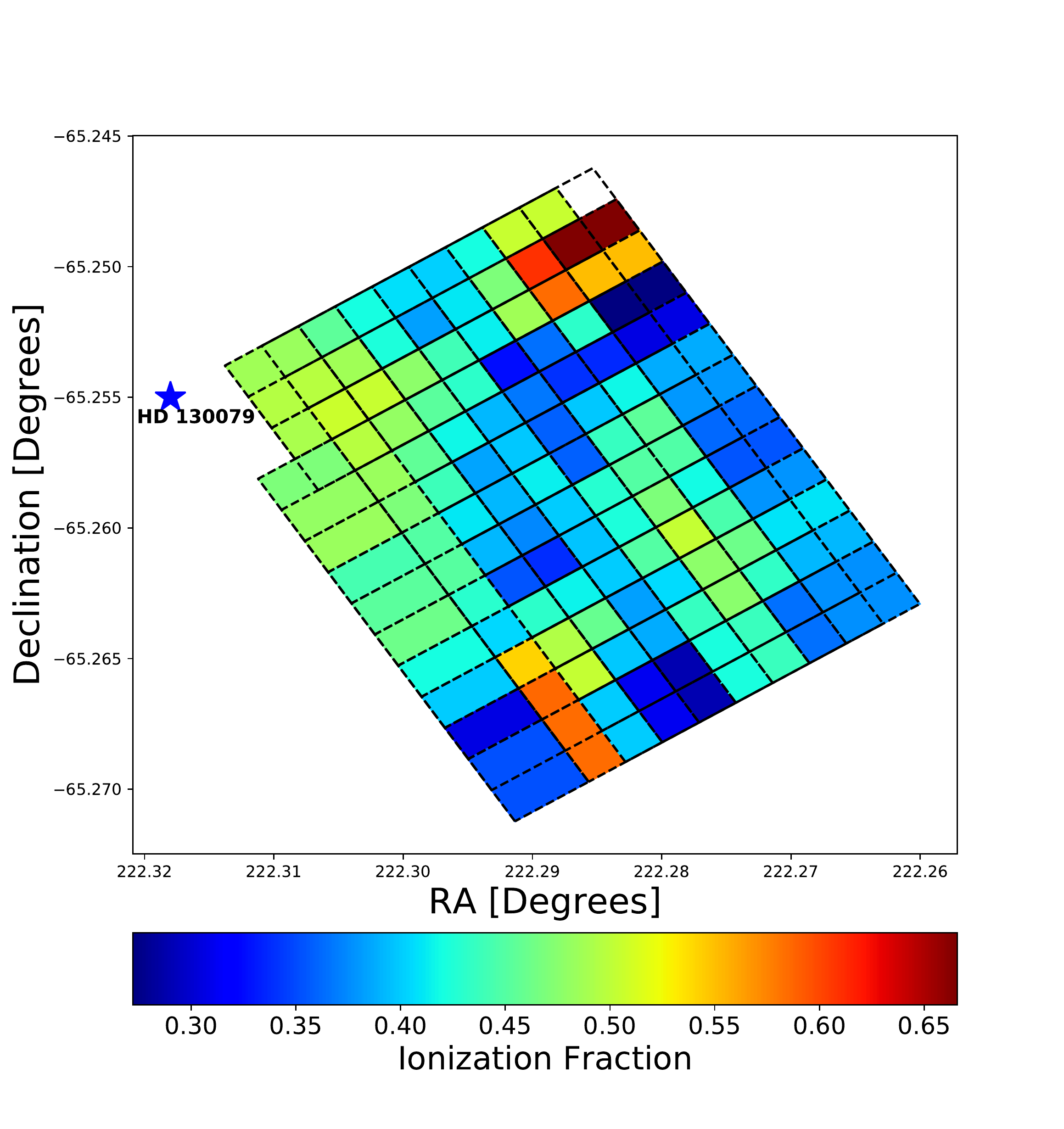} 
	\includegraphics[width=0.49\textwidth]{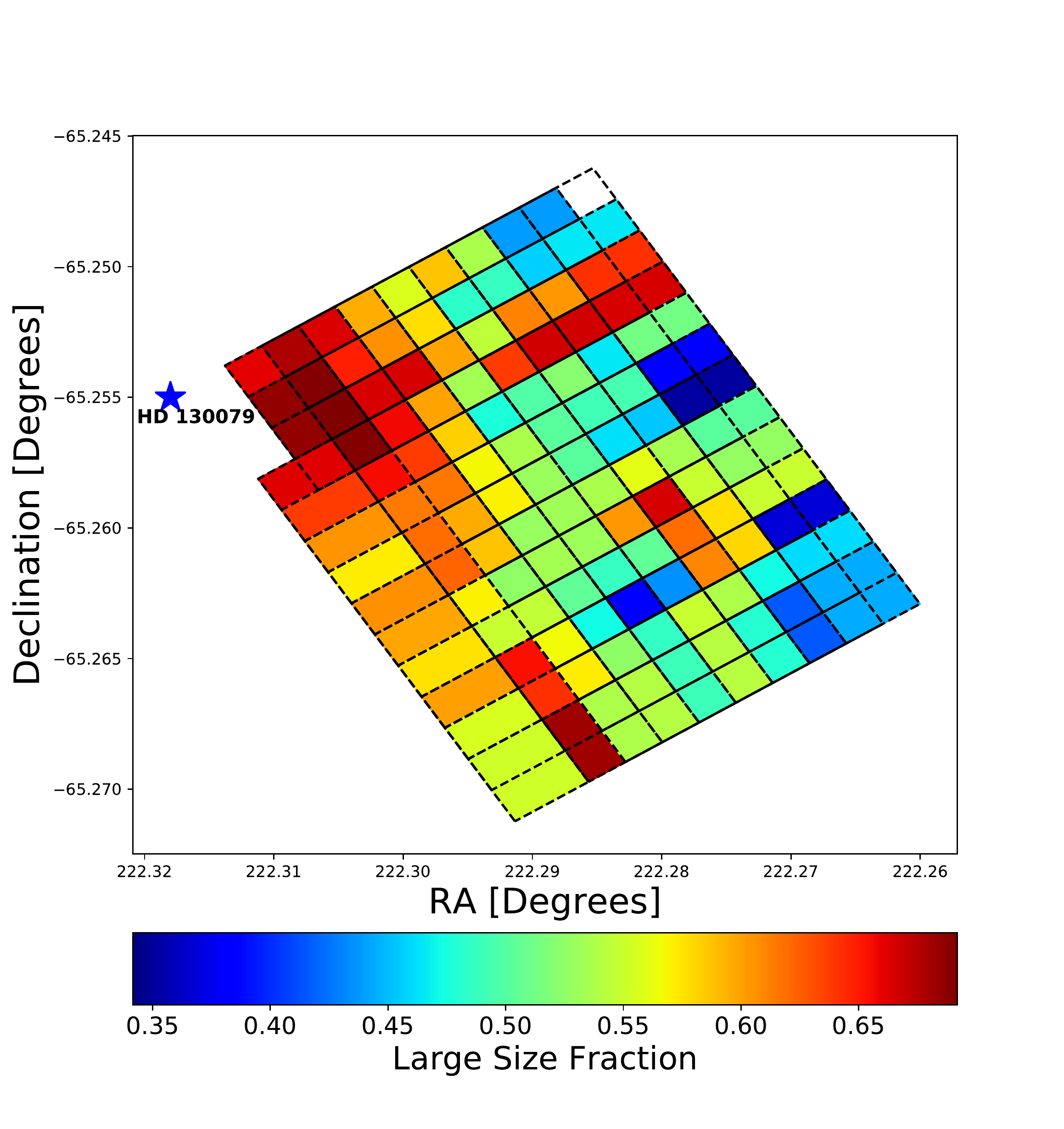} 
	\caption{(top left panel) Pypahdb values for the ionization fraction vs. distance with color coding according to the large size fraction, where large molecules are defined as $N_{\rm C}>40$). (top right panel) Pypahdb values for the ionization fraction vs. large size fraction with color coding according to the distance from HD\,130079. Data points included are those from regions with a SN\,$\ge 5$. (bottom) RA/DEC maps displaying the intensity of (left) the ionization fraction and (right) the large size fraction for the sampled regions.}
\label{ion_size}
\end{figure*}

For the rough estimates regarding the cation-to-neutral PAH ratio in DC\,314.8--5.1 subjected to the ionizing photon field of HD 130079, we therefore made the following approximations.  First, we assume the dominant \emph{gas} temperature component of $T\sim 800$\,K, according to the discussion by \cite{Boersma18} for the photo-dissociation region (PDR) in NGC\,7023. For the free electron number density, on the other hand, we follow the ``diffuse ISM approximation'', stating that all free electrons come from CII with the abundance of $1.4 \times 10^{-4}$ \citep[see][and references therein]{Boersma18}. Furthermore, we assume a standard power-law density profile for the globule, $n_{\rm H} \simeq n_0 \times (r/R)^{-1.3}$, where $r$ is the distance from the cloud center, and $R = \sqrt{a b}$ is the effective radius of an elliptical cloud characterized by the major and minor semi-axes $a$ and $b$, respectively. This, along with the values $2a \simeq 0.9$\,pc, $2b \simeq 0.6$\,pc,  and the total mass of the cloud $\simeq 160 M_{\odot}$ \citep[][corrected for the 435\,pc distance]{Whittet07}, gives $n_0 \sim 7 \times 10^3$\,cm$^{-3}$, and hence $n_e \sim 0.75$\,cm$^{-3}$ around $r \sim 0.45$\,pc\,$\gtrsim R$.

Finally, for estimating the ionizing photon field for HD\,130079, we take a stellar temperature of $T_{\star}=10,600$\,K,  radius $R_{\star} = 2.7 \times R_{\odot}$, and assume a blackbody emission spectrum for the starlight
\begin{equation}
    B_{\nu} \propto \frac{{\nu}^{3}}{\exp\left(\frac{h\nu}{kT}\right) - 1} \, ,
\end{equation}
such that the total luminosity of the star is $L_{\star} = 4\pi R^2_{\star} \, {\sigma}_{SB} T^4$. This gives the FUV luminosity of
\begin{equation}
    L_{\rm FUV} = f_{\rm FUV} \times L_{\star} \simeq 3 \times 10^{34} \, {\rm erg\,s^{-1}} \, ,
\end{equation}
where the bolometric correction factor is
\begin{equation}
    f_{\rm FUV} \equiv \frac{\int_{\nu_1}^{\nu_2}d\nu \, B_{\nu}}{\int_{0}^{\infty}d\nu \, B_{\nu}} \simeq 0.1
\end{equation}
with $h\nu_1 = 6$\,eV and $h\nu_2=13.6$\,eV. Consequently, at a given distance from the star $d$, where the FUV flux of the star is $F_{\rm FUV} = L_{\rm FUV}/4\pi d^2$, the ionizing photon field in Habing units reads as
\begin{equation}
    G_{0} = \frac{F_{\rm FUV}}{1.6\times 10^{-3} \, {\rm erg\,cm^{-2}\,s^{-1}}} \sim 174
\end{equation}
for $d=0.03$\,pc, which is the distance to the nearest sampled region to the star in DC\,314.8--5.1. We note that $G_0$ drops to $\sim$\,4 at the most distant region analyzed, at a distance of $d \simeq 0.18$\,pc (with no extinction correction taken into account).

 All the above estimates therefore imply $n_{\rm PAH+}/n_{\rm PAH0} \sim 0.5$ around the region nearest to HD\,130079 in DC\,314.8--5.1 ($d=0.03$\,pc) for the general, illustrative, value of $N_{\rm C} \sim 100$, and progressively smaller levels away from the photo-ionizing star, HD\,130079. The resulting ``outskirts level'' value is, in fact, not far off from the PAH cations-to-neutrals ratio, $\gtrsim 1$, implied for the region by the observed 6.2/11.2 PAH intensity ratio, as well as the 8.6/11.2 line ratio, both considered by \citet{Boersma18} to be good proxies for the ionized fraction parameter \citep[see also][]{Zang19}. However, the discrepancy between these values increases toward the globule's central regions (up to $d=0.18$\,pc), for which the Equation\,\ref{eq:ratio} implies, formally, $n_{\rm PAH+}/n_{\rm PAH0} < 0.01$, while the observed PAH intensity requires this ratio to be of the order of a few.

To illustrate the above-mentioned inconsistency, in Figure\,\ref{Ionization} we plot the 6.2/11.2 and 8.6/11.2 ratios (left and right columns respectively), as functions of distance from the star; in the top panels, we superimpose for illustration the lines corresponding to the intensity ratios expected for given values of the ionization parameter $n_{\rm PAH+}/n_{\rm PAH0}$, based on the best-fit correlations by \citet{Boersma18}, namely $I_{6.2 \mu m}/I_{11.2 \mu m} = 0.90 +0.71 \times (n_{\rm PAH+}/n_{\rm PAH0})$, and $I_{8.6 \mu m}/I_{11.2 \mu m} = 0.33 +0.29 \times (n_{\rm PAH+}/n_{\rm PAH0})$.

As shown in Figure\,\ref{Ionization}, both the 6.2/11.2 and 8.6/11.2 ratios, remain relatively high throughout the globule, revealing similar trends, with the values increasing somewhat toward the central regions of the cloud (however with significant spread of data points). With the \citeauthor{Boersma18} scaling relation, this would imply an overall high ionization $n_{\rm PAH+}/n_{\rm PAH0} > 1$ within the entire reflection nebula.

The breakdown enabled by the pypahdb fitting, offers a direct view on the ionization fraction within DC\,314.8--5.1. As shown in Figure\,\ref{ion_size}, the ionized fraction returned by the fitting procedure, $n_{\rm PAH+}/(n_{\rm PAH+}+n_{\rm PAH0})$, is $\simeq 0.5$ within the regions of the cloud closest to the ionizing star, and decreasing somewhat down to $0.3-0.4$ at further distances, again with an increasing spread. These values therefore correspond to a ratio $n_{\rm PAH+}/n_{\rm PAH0} \simeq 1$ at the outskirts of the cloud, closer to the star, and down to $\sim 0.5-0.7$ at the furthest distances from the star probed in our analysis. Based on this, we therefore conclude that we do see an overall decrease in the ionization fraction with distance, which is however much more modest than expected based solely on the decrease in the ionizing continuum from the star, hinting at the potential role of alternative ionization factors at play, such as ionization due to cosmic-rays.

Interestingly, the pypahdb fitting also provides insight into the large size fraction of the PAH molecules, where large molecules are defined as those with $N_{\rm C}>40$. The corresponding results are again shown in Figure\,\ref{ion_size}, revealing that the PAH emission of the highly ionized regions at the cloud's boundaries, is dominated by large molecules (large size fraction $> 0.6$), which become less prevalent for the regions closer to the center of the cloud (large size fraction $< 0.6$).

We furthermore perform a basic correlation study between the two parameters, ionization and large size fractions, over all of the sampled regions with a SN\,$\ge 5$, shown in the top right panel in Figure\,\ref{ion_size}. We utilize both a Pearson's product-moment correlation and Kendall's rank correlation. We find a statistically significant correlation with both methods, showing a statistic of 0.56 and 0.45 and p-value of $1.2 \times 10^{-10}$ and $3.6 \times 10^{-12}$ for Pearson's and Kendall's respectively.

\begin{figure}[!th]
\centering
        \includegraphics[width=\columnwidth]{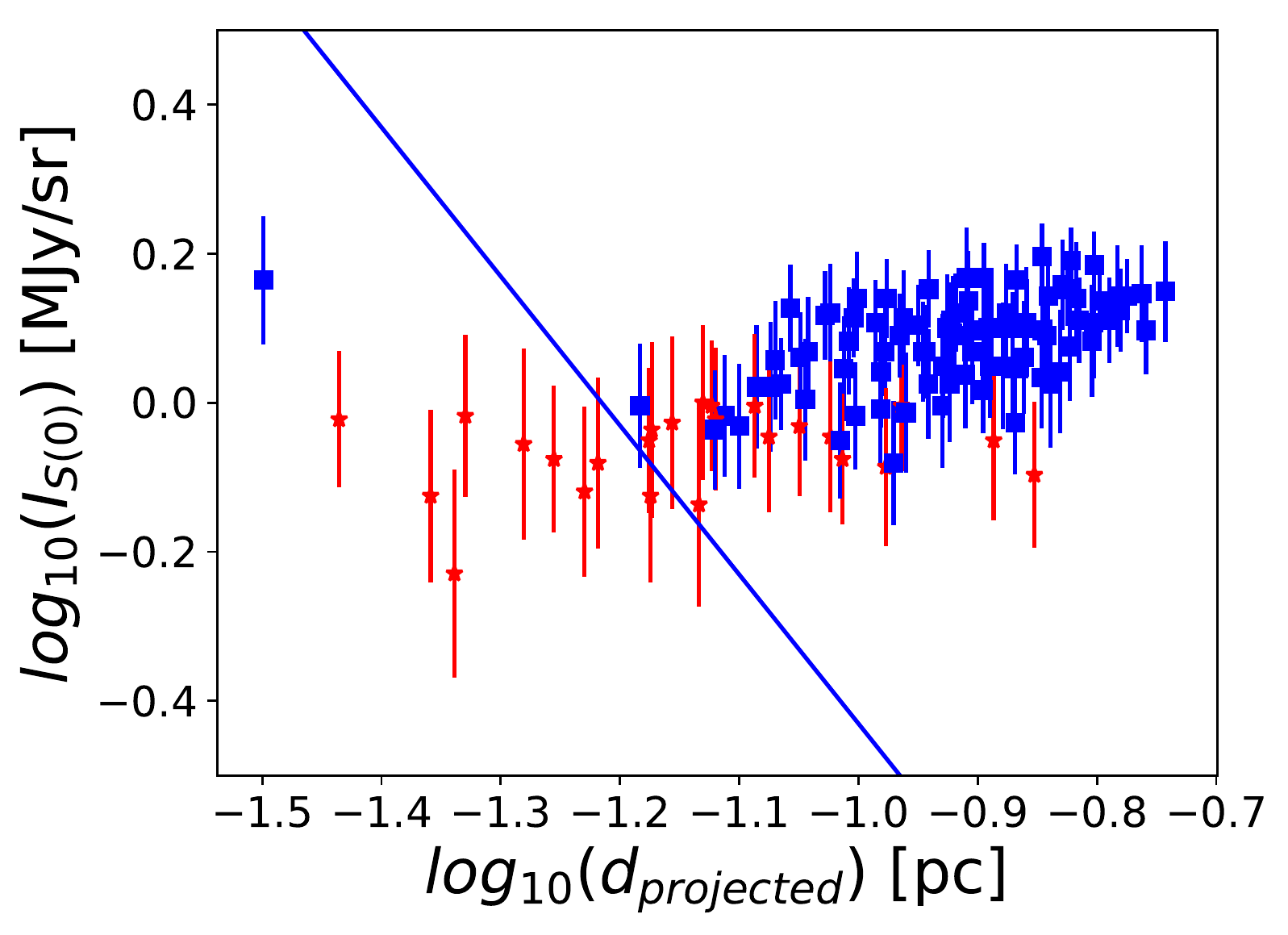}
        \includegraphics[width=\columnwidth]{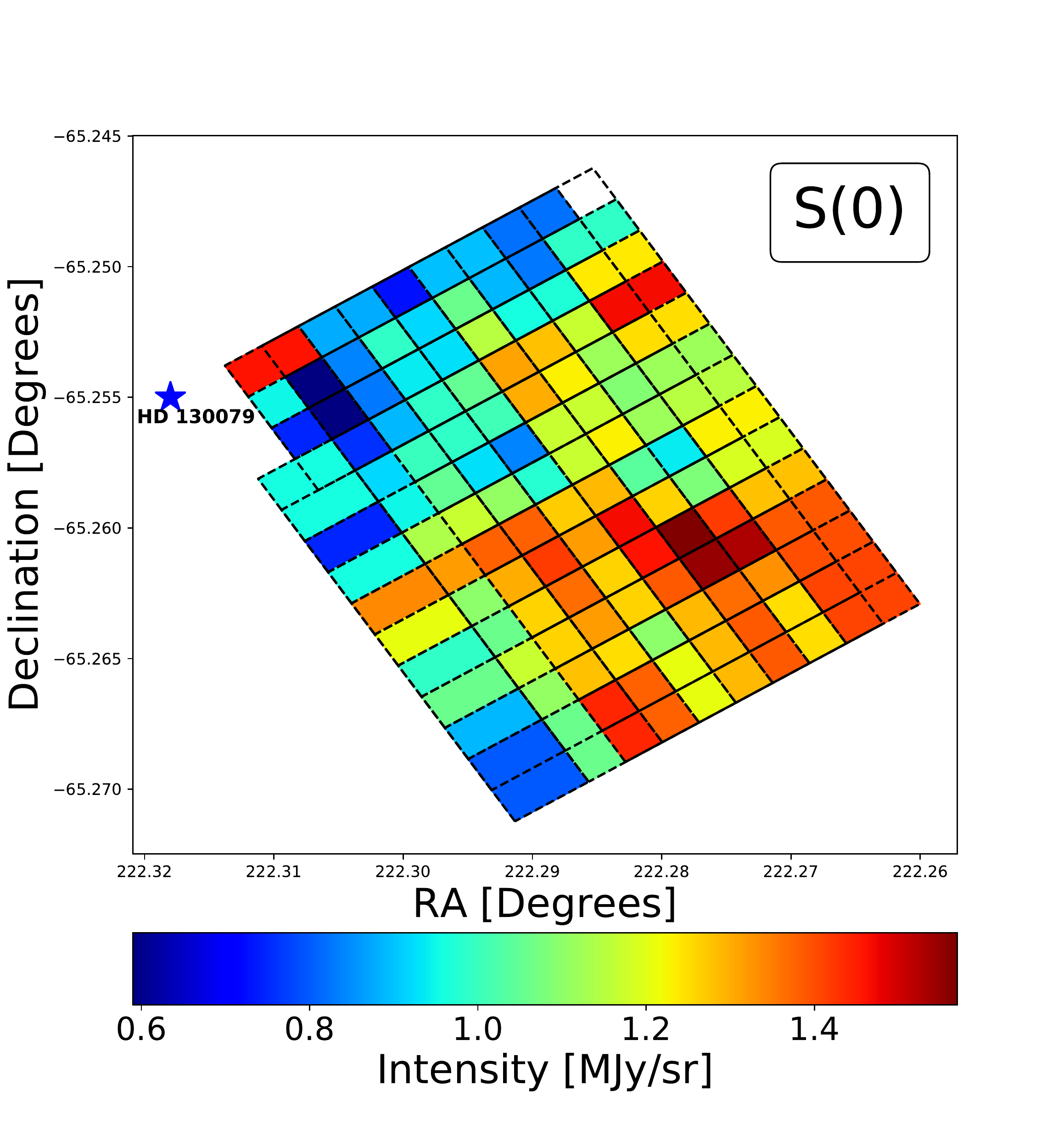}
    \caption{(top) Logarithmic intensity of the H$_2$\,S(0) line located at $28.1-28.2\,\mu$m plotted with distance from the field star, HD\,130079. The blue line represent the inverse square scaling (with arbitrary normalization), to guide the eye. SN is marked in each by color and symbol with red stars marking the lowest data points, $3<$\,SN\,$<5$, and blue squares marking all data points with SN\,$>5$.
    (bottom)  RA/DEC maps displaying the intensity of fitted H$_2$\,S(0) line in each the sampled regions with a SN\,$>3$.}
    \label{fig:hrot}
\end{figure}

\subsection{Molecular Hydrogen Detection}

As shown in Figure\,\ref{fig:hrot}, the H$_2$\,S(0) line is detected in our system at high significance (SN\,$>5$) only at larger distances from the star. On the other hand, its intensity remains fairly constant throughout those outer regions probed in our analysis.

The excitation for H$_2$ can occur through several processes, including inelastic collisions with gaseous species, radiative pumping from far-ultraviolet (FUV) radiation, or the formation process; the dominant source of the pure rotational line S(0) is collisional excitation, and as such the S(0) line intensity should be sensitive to the gas temperature at the PDR front \citep[e.g.,][]{LeBourlot99,Neufeld08,Habart11}. The H$_2$ excitation may however also result from interactions with secondary electrons generated by cosmic rays penetrating and ionizing the cloud \citep{Wakelam17}.

\section{Conclusions \& Final remarks}
\label{sec:conclusions}

In this paper, we have discussed the MIR spectroscopic properties, provided by the IRS instrument on the Spitzer Space Telescope, of the quiescent dark cloud, DC\,314.8--5.1. This study has focused on the lower-resolution MIR spectra in the range of 5--35\,$\mu$m. Spectra were extracted from 117 overlapping spatial regions over $\simeq 0.18$\,pc spanning the reflection nebula induced by the field star, HD 130079. Spectral fitting revealed a plethora of PAH features, the inspection of which led to the following conclusions:
\begin{enumerate}
\item The intensities of PAH features generally decrease over distance from the ionizing star toward the cloud center, some however showing a saturation (plateau) in the intensity profiles at larger distances.

\item  The relative intensities of both the 6.2 and 8.6 features with respect to the 11.2\,$\mu$m feature remain high throughout the globule, suggesting a relatively large cation-to-neutral PAH ratio $\gtrsim 1$; this value is consistent with the expected ionization level in the vicinity of the illuminating star, where the estimated ionization parameter $\chi \sim 10^4$; however, for the cloud's more central regions, where $\chi$ drops to $\sim 100$, a discrepancy emerges between the expected value and the one implied by the observed PAH intensity ratio.

\item The performed pypahdb fitting, confirms a high ionized fraction within the cloud, ranging from $\simeq 0.5$ within the regions in the closest vicinity to the ionizing star, down to $\sim 0.3-0.4$ at larger distances. Moreover, the PAH emission of the highly ionized regions at the cloud's boundaries, appears to be dominated by large molecules, which become less prevalent for the regions closer to the center of the cloud. 

\item  The investigation of the 7.7\,$\mu$m profile in the reflection nebula of DC\,314.8--5.1 suggests that, in addition to the chemically processed PAH population of the cloud, we also see a fraction of PAHs UV-processed by the the neighbouring star, HD\,130079.

\item We detected the H$_2$\,S(0) line at $28.1-28.2\,\mu$m with a higher significance at further distances from the star.

\end{enumerate}
All in all, our results hint at divergent physical conditions within the quiescent cloud DC\,314.8--5.1 as compared to molecular clouds with ongoing starformation, and may suggest a role played by cosmic-rays in the ionization of the system. 

\begin{acknowledgments}

This work was supported by the Fulbright Program and in collaboration with the Astronomical Observatory of the Jagiellonian University. E.K. and {\L}.S. were supported by Polish NSC grant 2016/22/E/ST9/00061. WRMR thanks the financial support from the Leiden Observatory. AK acknowledges support from the First TEAM grant of the Foundation for Polish Science No.\ POIR.04.04.00-00-5D21/18-00. This article has been supported by the Polish National Agency for Academic Exchange under Grant No.\ PPI/APM/2018/1/00036/U/001.

The authors thank A. W{\'o}jtowicz and D.C.B. Whittet for their discussion and comments.  The authors also thank the anonymous referee for their critical remarks and constructive suggestions, which helped to improved the quality of the paper.

This work is based on observations made with the Spitzer Space Telescope, obtained from the NASA/ IPAC Infrared Science Archive, both of which are operated by the Jet Propulsion Laboratory, California Institute of Technology under a contract with the National Aeronautics and Space Administration. 

This work has made use of data from the European Space Agency (ESA) mission Gaia (\url{https://www.cosmos.esa.int/gaia}), processed by the Gaia Data Processing and Analysis Consortium (DPAC, \url{https://www.cosmos.esa.int/web/gaia/dpac/consortium}). Funding for the DPAC has been provided by national institutions, in particular the institutions participating in the Gaia Multilateral Agreement. We are grateful to Timo Prusti for advice on Gaia data.

\end{acknowledgments}

\end{document}